\pdfoutput=1
\documentclass[cernpreprint, atlasdraft=false, UKenglish, texmf, orcidlogo]{atlasdoc} 
 
\usepackage{atlaspackage}
\usepackage{atlasbiblatex}
 
\usepackage{atlasphysics}
\usepackage{feynmp-auto}

\graphicspath{{logos/}{figures/}}
 
\usepackage{ANA-STDM-2020-02-PAPER-defs}

\AtlasTitle{Differential cross-section measurements
of the production of four charged leptons in association with two jets
using the ATLAS detector}

\AtlasAbstract{
Differential cross-sections are measured for the production of four charged leptons in association with two jets.
These measurements are sensitive to
final states in which the jets are produced via the strong interaction as well as to the purely-electroweak vector boson scattering process.
The analysis is performed using proton--proton collision data collected by ATLAS at $\sqrt{s}=13$~TeV and with an integrated luminosity of 140~fb$^{-1}$. The data are corrected for the effects of detector inefficiency and resolution and
are compared to state-of-the-art Monte Carlo event generator predictions.
The differential cross-sections are used to search for anomalous weak-boson self-interactions that are induced by dimension-six and dimension-eight operators in Standard Model effective field theory.
}

\AtlasRefCode{STDM-2020-02}
 
\PreprintIdNumber{CERN-EP-2023-152}

\AtlasJournal{JHEP}
\AtlasJournalRef{\JHEP 01 (2024) 004}
\AtlasDOI{10.1007/JHEP01(2024)004}

\hypersetup{pdftitle={ATLAS document},pdfauthor={The ATLAS Collaboration}}
 
\begin{document}
 
\maketitle
 
\tableofcontents

\section{Introduction}
\label{sec:intro}

\newcommand{\AtlasCoordFootnote}{
ATLAS uses a right-handed coordinate system with its origin at the nominal interaction point (IP)
in the centre of the detector and the \(z\)-axis along the beam pipe.
The \(x\)-axis points from the IP to the centre of the LHC ring,
and the \(y\)-axis points upwards.
Cylindrical coordinates \((r,\phi)\) are used in the transverse plane,
\(\phi\) being the azimuthal angle around the \(z\)-axis.
The pseudorapidity is defined in terms of the polar angle \(\theta\) as \(\eta = -\ln \tan(\theta/2)\).
Angular distance is measured in units of \(\Delta R \equiv \sqrt{(\Delta\eta)^{2} + (\Delta\phi)^{2}}\).}
 
\begin{figure}[tp]
\centering
\includegraphics[width=0.65\textwidth]{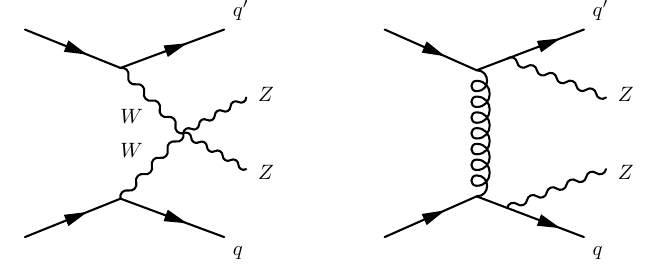}
\caption{Example Feynman diagrams for EW \zzjjtwo{} production (left) and strong \zzjjtwo{} production (right). The scattered quarks (labelled $q$ and $q^\prime$) produce the hadronic jets observed in the final state. \label{fig:zzjj}}
\end{figure}

The pair production of $Z$ bosons in association with two jets (\zzjjtwo{} production) in proton--proton collisions is sensitive to a diverse range of physical phenomena. Of particular interest is the purely electroweak (EW) \zzjjtwo{} process (referred to here as EW \zzjjtwo{})  in which the jets arise through the $t$-channel exchange of electroweak bosons as shown in Figure~\ref{fig:zzjj} (left). This process is sensitive to the  $WWZ$ and $WWZZ$ weak-boson self-interactions, which arise due to the non-Abelian nature of the electroweak interaction. Also of interest is the production of \zzjjtwo{} with the jets arising from the strong interaction. This is referred to as strong \zzjjtwo{} production and is shown in Figure~\ref{fig:zzjj} (right). Theoretical predictions for the strong \zzjjtwo{} process are very sensitive to the accuracy of the perturbative QCD calculations,
both in the overall production rate and in the kinematic properties of the final state. Measurements of \zzjjtwo{} production can therefore be used to improve our understanding of both the electroweak and strong interactions that underpin the Standard Model (SM) of particle physics.
 
Measurements of \zzjjtwo{} production in proton--proton collisions at the LHC have thus far focussed on extracting the electroweak contribution. The ATLAS Collaboration recently published the observation of EW $ZZjj$ production at a significance of greater than five standard deviations~\cite{ATLAS:2020nlt}. Before that, the CMS Collaboration  announced evidence for the process at a significance greater than three standard deviations~\cite{CMS:2020fqz}. Both of these measurements used multivariate discriminants to separate the electroweak and strong \zzjjtwo{} components, and reported a fiducial EW \zzjjtwo{} cross-section within the phase space of the analysis.
In the ATLAS analysis, a cross-section for inclusive \zzjjtwo{} production was also reported. The inclusive \zzjjtwo{} cross-section is the sum of the strong and electroweak contributions, and the ATLAS measurement provides an important constraint on the overall rate of strong \zzjjtwo{} production. Measurements sensitive to inclusive $ZZ$ production have also been performed by both experiments, in an inclusive phase space that is not enriched by EW production~\cite{CMS:2020gtj,ATLAS:2021kog}.

In this paper, differential cross-section measurements for the production of four charged leptons in association with two jets are reported (\zzjj{} production, with $\ell=e,\mu$). For four-lepton invariant masses greater than twice the $Z$ boson mass, this final state is dominated by the strong \zzjjtwo{} and EW \zzjjtwo{} processes, with the $Z$ bosons decaying via $ZZ\rightarrow \ell^+ \ell^- \ell^+ \ell^-$. Contributions from $\gamma^*$ and  $Z^*$ are increasingly important at four-lepton invariant masses that are lower than twice the $Z$ boson mass. Differential cross-sections are particularly useful for probing the electroweak and strong production mechanisms. In the case of EW \zzjj{} production, differential cross-sections can be used to search for anomalous $WWZ$ and $WWZZ$ interactions, which have different Lorentz structures to the corresponding interactions in the SM and can therefore distort the kinematic properties of the electroweak process. For strong \zzjj{} production, the differential cross-sections allow the perturbative QCD calculations to be confronted in extreme phase space regions such as high dijet invariant mass. The differential cross-sections are measured as a function of three types of observables:
\begin{itemize}
\item \textit{Vector boson scattering (VBS) observables} are those that are used to  characterise the vector-boson scattering  process shown in Figure \ref{fig:zzjj} (left). The measured observables are the four-lepton invariant mass, \mfl{}, the transverse momentum of the four-lepton system, \pTfl{}, the dijet invariant mass, \mjj{}, the rapidity interval spanned by the dijet system, \Dyjj{}, and the transverse momentum of the dijet system, \pTjj{}.\footnote{\AtlasCoordFootnote}
\item \textit{Polarisation, charge conjugation, and parity observables}: The polarisation of the $Z$ boson can be probed using the cosine of the angle between the negatively-charged lepton and the $Z$ boson as measured in the centre-of-mass frame of the $Z$ boson. Two angles are measured, \cosTS{} and \cosTStwo{}, which are sensitive to the polarisation of the leading and subleading $Z$ boson candidates, respectively (the leading $Z$ boson candidate is defined as part of the event selection in Section~\ref{sec:evsel}). The charge conjugation (C) and parity (P) structure of the $WWZ$ and $WWZZ$ interactions can be probed using the signed azimuthal angle between the two jets, $\Dphijj = \phi_f -\phi_b$, where the jets in the dijet system are ordered such that $y_f>y_b$~\cite{Hankele:2006ma}.
\item \textit{QCD-sensitive observables} probe higher-order real emission of quarks and gluons from the \zzjj{} processes. The measured observables are the transverse momentum of the \zzjj{} system, \pTfljj{}, and the scalar sum of transverse momentum of the four leptons and the two jets, \STfljj{}.
\end{itemize}
The observables are measured in VBS-enhanced and VBS-suppressed phase space regions. The selection cuts that define the fiducial region of the differential cross-section measurements is given in Section~\ref{sec:unfolding}.
 
The differential cross-sections are sensitive to anomalous weak-boson self-interactions, which are predicted by effective field theory (EFT)  extensions to the SM. Of particular interest are the operators in dimension-eight EFT, which induce anomalous quartic interactions (such as $WWZZ$) without any  corresponding triple gauge coupling (such as $WWZ$) ~\cite{Eboli:2016kko}. Such operators can only be tested using VBS and triboson processes. The differential cross-sections measured in this paper are used in Section~\ref{sec:eft} to constrain the contribution of these purely quartic weak-boson self-interactions.


\section{ATLAS detector}
\label{sec:detector}


The ATLAS detector~\cite{PERF-2007-01} at the LHC covers nearly the entire solid angle around the collision point.
It consists of an inner tracking detector surrounded by a thin superconducting solenoid, electromagnetic and hadron calorimeters,
and a muon spectrometer incorporating three large superconducting air-core toroidal magnets.
 
The inner-detector system (ID) is immersed in a \SI{2}{\tesla} axial magnetic field
and provides charged-particle tracking in the range \(|\eta| < 2.5\).
The high-granularity silicon pixel detector covers the vertex region and typically provides four measurements per track,
the first hit normally being in the insertable B-layer (IBL) installed before Run~2~\cite{ATLAS-TDR-19,PIX-2018-001}.
It is followed by the silicon microstrip tracker (SCT), which usually provides eight measurements per track.
These silicon detectors are complemented by the transition radiation tracker (TRT),
which enables radially extended track reconstruction up to \(|\eta| = 2.0\).
The TRT also provides electron identification information
based on the fraction of hits (typically 30 in total) above a higher energy-deposit threshold corresponding to transition radiation.
 
The calorimeter system covers the pseudorapidity range \(|\eta| < 4.9\).
Within the region \(|\eta|< 3.2\), electromagnetic calorimetry is provided by barrel and
endcap high-granularity lead/liquid-argon (LAr) calorimeters,
with an additional thin LAr presampler covering \(|\eta| < 1.8\)
to correct for energy loss in material upstream of the calorimeters.
Hadron calorimetry is provided by the steel/scintillator-tile calorimeter,
segmented into three barrel structures within \(|\eta| < 1.7\), and two copper/LAr hadron endcap calorimeters.
The solid angle coverage is completed with forward copper/LAr and tungsten/LAr calorimeter modules
optimised for electromagnetic and hadronic energy measurements respectively.
 
The muon spectrometer (MS) comprises separate trigger and
high-precision tracking chambers measuring the deflection of muons in a magnetic field generated by the superconducting air-core toroidal magnets.
The field integral of the toroids ranges between \num{2.0} and \SI{6.0}{\tesla\metre}
across most of the detector.
Three layers of precision chambers, each consisting of layers of monitored drift tubes, cover the region \(|\eta| < 2.7\),
complemented by cathode-strip chambers in the forward region, where the background is highest.
The muon trigger system covers the range \(|\eta| < 2.4\) with resistive-plate chambers in the barrel, and thin-gap chambers in the endcap regions.
 
Interesting events are selected by the first-level trigger system implemented in custom hardware,
followed by selections made by algorithms implemented in software in the high-level trigger~\cite{TRIG-2016-01}.
The first-level trigger accepts events from the \SI{40}{\MHz} bunch crossings at a rate below \SI{100}{\kHz},
which the high-level trigger further reduces in order to record events to disk at about \SI{1}{\kHz}.
 
An extensive software suite~\cite{ATL-SOFT-PUB-2021-001} is used in the reconstruction and analysis of real
and simulated data, in detector operations, and in the trigger and data acquisition systems of the experiment.


\section{Data and simulation}
\label{sec:datasimul}

 
The data used in this measurement correspond to an integrated luminosity of 140~\ifb and
were recorded using single-lepton and multi-lepton triggers. The minimum trigger thresholds on transverse momentum depends on the lepton flavour, lepton multiplicity and data-taking periods, and vary between $20-26$ \GeV~and $8-24$ \GeV~for the single-lepton and multi-lepton triggers, respectively. The leptons that fire the trigger are required to be part of the four-lepton system that defines the signal region of the analysis (discussed in Section~\ref{sec:evsel}). The trigger efficiency for inclusive \zzjj{} events is close to unity.
 
The signal and background processes were modelled using Monte Carlo (MC) event generator simulations. Strong \zzjj{} production was modelled using \sherpa{}~2.2.2~\cite{Bothmann:2019yzt}.
The fully leptonic final state was simulated using
matrix elements at next-to-leading-order (NLO) accuracy in perturbative QCD for up to one additional parton
and at leading-order (LO) accuracy for up to three additional parton
emissions. Samples for the loop-induced process \(gg \to ZZ\) were
simulated using matrix elements accurate a leading order  (LO) in QCD for up to one
additional parton emission, normalised using an \mfl{}-dependent $k$-factor at NLO accuracy in QCD~\cite{Caola:2015psa,Caola:2016trd}, and including an additional, constant correction of about 1.2
to reproduce next-to-next-to-leading-order (NNLO) calculations for the Higgs-mediated process~\cite{Passarino:2013bha,LHCHiggsCrossSectionWorkingGroup:2016ypw}.
Despite of the approximation of NNLO effects, the normalisation uncertainty for the \(gg \to ZZ\) process is still conservatively considered from the NLO calculation.
The matrix element calculations were matched
and merged with the \sherpa{} parton shower based on Catani--Seymour
dipole factorisation~\cite{Gleisberg:2008fv,Schumann:2007mg} using the MEPS@NLO
prescription~\cite{Hoeche:2011fd,Hoeche:2012yf,Catani:2001cc,Hoeche:2009rj}.
The virtual QCD corrections were provided by the
\textsc{OpenLoops} library~\cite{Buccioni:2019sur,Cascioli:2011va,Denner:2016kdg}. The
\textsc{NNPDF3.0nnlo} set of parton distribution functions (PDF) were used~\cite{Ball:2014uwa}, along with the
dedicated set of tuned parton-shower parameters developed by the
\sherpa{} authors.  This prediction is referred to as \sherpa{} strong \zzjj{}.
 
An alternative prediction for strong \zzjj{} production is obtained using \fxfx{}~\cite{Alwall:2014hca}. The sample was simulated using
matrix elements at NLO accuracy in QCD for up to one parton in the final state, and with the PDF4LHC15 NLO set of PDFs~\cite{Butterworth:2015oua}. The \fxfx{} generator was interfaced to \pythia{} to provide parton showering, hadronisation, and underlying-event activity, using the A14 set of tuned parameters~\cite{ATL-PHYS-PUB-2014-021}. To remove overlap between the matrix element and the parton shower, the different jet multiplicities were merged using the FxFx prescription~\cite{Frederix:2012ps}. \evtgen{} was used for the properties of the bottom and charm hadron decays. The \(gg \to ZZ\) contribution is again estimated by using \sherpa{}. This prediction is referred to as \mgnlo{} strong \zzjj{}.
 
Electroweak \zzjj{} production is modelled using \mg{}~\cite{Alwall:2014hca}. The Feynman diagrams used in the calculation include the $t$-channel exchange of an electroweak boson and the  $s$-channel processes that contribute at the same order in $\alpha_{\textrm{EW}}^{}$, i.e  $ZZV$ production (with $V\rightarrow jj$). The calculation is accurate to LO in perturbative QCD and uses the \textsc{NNPDF3.0nlo} PDF set. The events were passed through \pythia{} to provide parton showering, hadronisation, and underlying-event activity using the A14 set of tuned parameters. This prediction is referred to as \mglo{} EW \zzjj{}.
 
An alternative prediction for EW \zzjj{}  production is obtained using the \textsc{Powheg-Box}~v2 event generator~\cite{Frixione:2007nw,Nason:2004rx,Frixione:2007vw,Alioli:2010xd,Jager:2013iza}. The Feynman diagrams are restricted to the $t$-channel exchange of an electroweak boson and the matrix elements calculated at NLO in perturbative QCD, using the \textsc{NNPDF3.0nlo} parton distribution functions. The events are passed through \pythia{} to produce the fully hadronic final state, using the A14 set of tuned parameters.  The $s$-channel contributions from $ZZV$ production are estimated  with the
\sherpa{}~2.2.2 generator, at NLO accuracy for the inclusive process and to LO accuracy for up to
two additional parton emissions using the \textsc{NNPDF3.0nnlo} PDF set and the dedicated set of tuned parton-shower parameters developed by the \sherpa{} authors. The triboson events are added to those from \powpyt{} to produce the alternative EW \zzjj{} prediction. The triboson contribution is about 7\% in the fiducial region used in this measurement. This prediction is referred to as  \powpyt{} EW \zzjj{}.
 
Background processes with four prompt leptons arise from $t\bar{t}Z$ production as well as $WWZ$ and $WZZ$ production.  The production of $t\bar{t}Z$ events was modelled using the \sherpa{}~2.2.0 generator at LO accuracy, using
the MEPS@LO set-up~\cite{Catani:2001cc,Hoeche:2009rj} with up to one
additional parton. The default
\sherpa{} parton shower was used along with the
\textsc{NNPDF3.0nnlo}~\cite{Ball:2014uwa} PDF set. The sample is scaled such that it reproduces the ATLAS measurement of $t\bar{t}Z$ production~\cite{ATLAS:2019fwo}. The production of $WWZ$ and $WZZ$ events were simulated with the
\sherpa{}~2.2.2 generator. The predictions are accurate to NLO in QCD for the inclusive process and to LO for up to
two additional parton emissions. The \textsc{NNPDF3.0nnlo} PDF set was used along with the dedicated set of tuned parton-shower parameters developed by the \sherpa{} authors. The sample is scaled such that it reproduces the ATLAS measurement of triboson production~\cite{ATLAS:2019dny}.

Background processes that do not contain four prompt leptons are estimated by using a data-driven technique as outlined in Section~\ref{sec:background}. The method is cross-checked using simulations of the relevant processes. Production of $WZjj$ with the subsequent leptonic decays of vector bosons was modelled with \textsc{Sherpa}~2.2.2, using the same approach as in strong \zzjj{} production. Non-prompt backgrounds arising from $Z+\textrm{jets}$ were simulated using \sherpa{}~2.2.1~\cite{Bothmann:2019yzt}
at NLO accuracy in perturbative QCD for up to two partons and LO accuracy
for up to four partons. The matrix elements were calculated with the Comix~\cite{Gleisberg:2008fv}
and \textsc{OpenLoops} libraries and
matched with the \sherpa{} parton shower using the MEPS@NLO
prescription. The sample was produced with the \textsc{NNPDF3.0nnlo} PDF set, the dedicated set of tuned parton-shower parameters developed by the \sherpa{} authors, and was normalised to a prediction accurate to NNLO
in QCD~\cite{Anastasiou:2003ds}. The production of \ttbar events was modelled using the
\powbox{}~v2~\cite{Frixione:2007nw,Nason:2004rx,Frixione:2007vw,Alioli:2010xd}
generator at NLO with the \textsc{NNPDF3.0nlo} PDF set
and the \hdamp parameter set
to 1.5 times the top mass~\cite{ATL-PHYS-PUB-2016-020}. The events were interfaced
to \pythia{} to model the parton shower,
hadronisation, and underlying event, with parameters set according
to the A14 tune.

All samples were passed through a detailed simulation of the ATLAS detector~\cite{SOFT-2010-01} based on \textsc{Geant4}~\cite{Agostinelli:2002hh}. Simulated inelastic $pp$ collisions were overlaid to model additional $pp$ collisions in the same and neighbouring bunch crossings (pile-up). 
The simulated events were then reweighted to match the pile-up conditions in the data. All simulated events were processed using the same reconstruction algorithms as used in data.
Furthermore, the lepton and jet momentum scale and resolution, and the lepton reconstruction, identification, isolation and trigger efficiencies in the simulation were corrected to match those measured in data.


\label{sec:simu}
 
\section{Event and object selection}

 
Events are selected for analysis if they were recorded during stable beam conditions and if they satisfy stringent data-quality requirements \cite{DAPR-2018-01}.
Proton--proton interaction vertices are reconstructed using ID tracking information~\cite{PERF-2015-01,} and each reconstructed vertex is required to have at least two associated tracks. Events are required to have at least one reconstructed vertex, with the primary vertex defined as the one with the largest sum of squared track transverse momenta.
 
Muons are reconstructed from information in the MS and the ID. \textit{Baseline} muons are required to satisfy the `Loose' identification criteria \cite{MUON-2018-03} and are required to be associated with the primary hard-scatter vertex by requiring $|z_0 \textrm{sin} \theta| < 0.5$~mm, where $z_0$ is the longitudinal difference between the primary vertex and the point at which the muon transverse impact parameter is measured. Baseline muons are required to have $p_{\textrm T}>5$~\GeV and $|\eta|<2.7$. \textit{Signal} muons are required to satisfy the baseline muon criteria and the `Loose' particle-flow-based isolation working point~\cite{MUON-2018-03}. They are also required to satisfy $d_0/\sigma_{d_0} < 3$, where $d_0$ is the transverse impact parameter calculated relative to the measured beam-line position and $\sigma_{d_0}$ is its uncertainty.
 
Electrons are reconstructed from topological clusters of energy deposited in the electromagnetic calorimeter that are matched to an ID track.
\textit{Baseline} electrons are required to satisfy the `VeryLoose' identification criteria \cite{PERF-2017-01} and to be associated with the primary hard-scatter vertex, by requiring $|z_0 \textrm{sin} \theta| < 0.5$~mm. 
Baseline electrons are required to have $p_{\textrm T}>7$~\GeV and $|\eta|<2.47$. \textit{Signal} electrons are required to satisfy the baseline electron criteria and the `LooseAndBLayer' identification \cite{PERF-2017-01} and `Loose' isolation \cite{EGAM-2018-01} working points. They are also required to satisfy  $d_0/\sigma_{d_0} < 5$.
 
Jets are reconstructed using the anti-$k_t$ algorithm \cite{Cacciari:2008gp,Cacciari:2011ma} with a radius parameter of $R=0.4$. The inputs to the algorithm are objects constructed using the particle-flow algorithm \cite{PERF-2015-09}, based on noise-suppressed positive-energy topological clusters in the calorimeter. Energy deposited in the calorimeter by charged particles is subtracted and replaced by the momenta of ID tracks which are matched to those topological clusters. The jets are initially calibrated using simulations and corrected using in situ measurements of the jet energy scale that is determined from dijet, $\gamma+\textrm{jet}$ and $Z+\textrm{jet}$ events \cite{JETM-2018-05}. Jets are required to have $p_{\textrm T}>30$~\GeV and $|\eta|<4.5$. To reduce the impact of jets that originate from pile-up interactions, jets with $|\eta|<2.4$ and $p_{\textrm T}<60$~\GeV, or with $2.4<|\eta|<4.5$ and $p_{\textrm T}<50$~\GeV, are required to satisfy the `Tight' working points of the jet vertex tagging algorithms~\cite{PERF-2014-03, PERF-2016-06}. To remove leptons reconstructed as jets, any jets within the range $\Delta R <0.2$ of an electron are rejected. A similar requirement is applied to jets that overlap with muons, if there are less than three ghost-associated~\cite{Cacciari:2008gn} ID tracks within the jet.
 
Events are required to have at least four (baseline) leptons. The two leptons with the largest transverse momentum are required to satisfy $p_{\textrm T}>20$~\GeV. All possible combinations of same-flavour opposite-charge (SFOC) lepton pairs are formed and each pair is required to satisfy $m_{\ell \ell}>5$~\GeV and $\Delta R(\ell,\ell)>0.05$, which reduces backgrounds from the leptonic decays of hadrons. The SFOC pairs are then ordered by $|m_{\ell \ell} - m_Z|$. The two $Z$-boson candidates are defined as the two SFOC pairs that have the smallest value of $|m_{\ell \ell} - m_Z|$ and are formed from different leptons. The leading $Z$-boson candidate is defined as the one that has the largest value of $|y_{\ell \ell}|$. The invariant mass of the four leptons is required to satisfy $\mfl>130$~\GeV and each lepton in the quadruplet is required to satisfy the signal lepton definition discussed earlier. Events are also required to contain at least two jets, with the highest transverse momentum jet satisfying $p_{\textrm T}>40$~\GeV. The dijet system is then defined as the two leading (highest transverse momentum) jets in the event that have $\eta_{j_1} \times \eta_{j_2}<0$. The dijet system is required to satisfy  $\Dyjj>2.0$ and $\mjj>300$~\GeV.
 
The events that satisfy the  selections listed above are then divided into \textit{VBS-enhanced} and \textit{VBS-suppressed} regions using the centrality of the four-lepton system,
\begin{equation}
\label{eq:cent}
\zeta = \left|\frac{\left[y_{4\ell} - 0.5(y_{j_1}+y_{j_2})\right]}{\Delta y_{jj}} \right|,
\end{equation}
where $y_{4\ell}$ is the rapidity of the four lepton system and $y_{j_1}$ ($y_{j_2}$) is the rapidity of the leading (subleading) jet in the dijet system. The VBS-enhanced (VBS-suppressed) region is defined as $\zeta<0.4$ ($\zeta>0.4$).
 
The measured and predicted event yields in the VBS-enhanced and VBS-suppressed regions are shown in Table~\ref{tab:eventyield}.
The event yields as a function of \mjj{} and \mfl{} are shown in Figure~\ref{fig:dataMCcomparison}, for both the VBS-enhanced and VBS-suppressed regions. In total, 169 events are reconstructed in the VBS-enhanced region and 53 events in the VBS-suppressed region. In both of the regions, the backgrounds are small and the event yields are dominated by \zzjj{} production. In the VBS-enhanced region, the electroweak process is predicted to be about 17\% of the event yield, but this rises to about 40\% of the event yield at high-\mjj{}. There is a modest contribution from background processes that is typically at the 5\%--10\% level, but reaching 20\% at the lowest values of \mfl{}. The background arising from non-prompt leptons is calculated using a data-driven technique as outlined in Section~\ref{sec:background}. Overall, the data are in good agreement with the signal simulations and predicted backgrounds.
 
\begin{table} [tp]
\centering
\begin{tabular}{l c c }
\hline \hline
Process &\multicolumn{2}{c}{Event yield $\pm$ stat. $\pm$ syst.}\\
\hline
& VBS-enhanced & VBS-suppressed \\
\hline
strong \zzjj{} (\sherpa)& $98.9  \pm 0.5 \pm 25.2$ & $45.5  \pm 0.3 \pm 12.9 $ \\
EW \zzjj{} (\mglo) & $24.1  \pm  0.1\ \pm  1.8$ & $2.12  \pm  0.02 \pm  0.14 $ \\
Prompt background & $18.8  \pm 0.2\ \pm 2.2$ & $5.5  \pm 0.1 \pm 0.4  $ \\
Non-prompt background & $3.0  \pm 0.6 \pm 3.2$ & $1.1  \pm 0.5 \pm 1.2 $ \\
\hline
Total prediction & $144  \pm1 \pm 26$ & $54  \pm 1 \pm 13 $ \\
Data & 169  &  53  \\
\hline
\end{tabular}
\vspace{0.5cm}
\caption{Measured and predicted event yields in the VBS-enhanced and VBS-suppressed regions. The background arising from non-prompt leptons is calculated using a data-driven technique as outlined in Section~\ref{sec:background}. The systematic uncertainties in the predictions are estimated by using the procedure outlined in  Section~\ref{sec:systematics}.}
\label{tab:eventyield}
\end{table}
 
\begin{figure}[tp]
\includegraphics[width=0.5\textwidth]{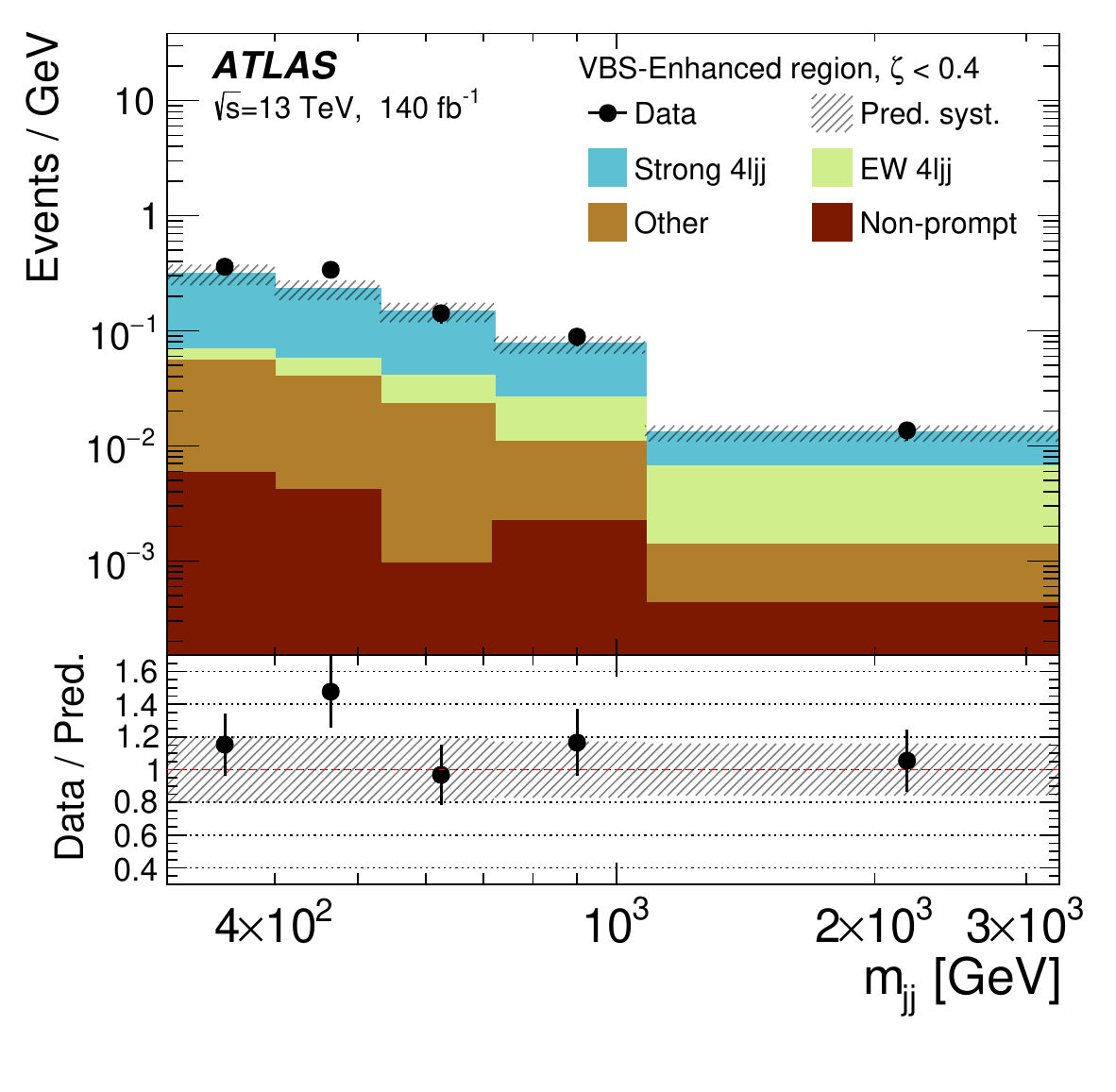}
\includegraphics[width=0.5\textwidth]{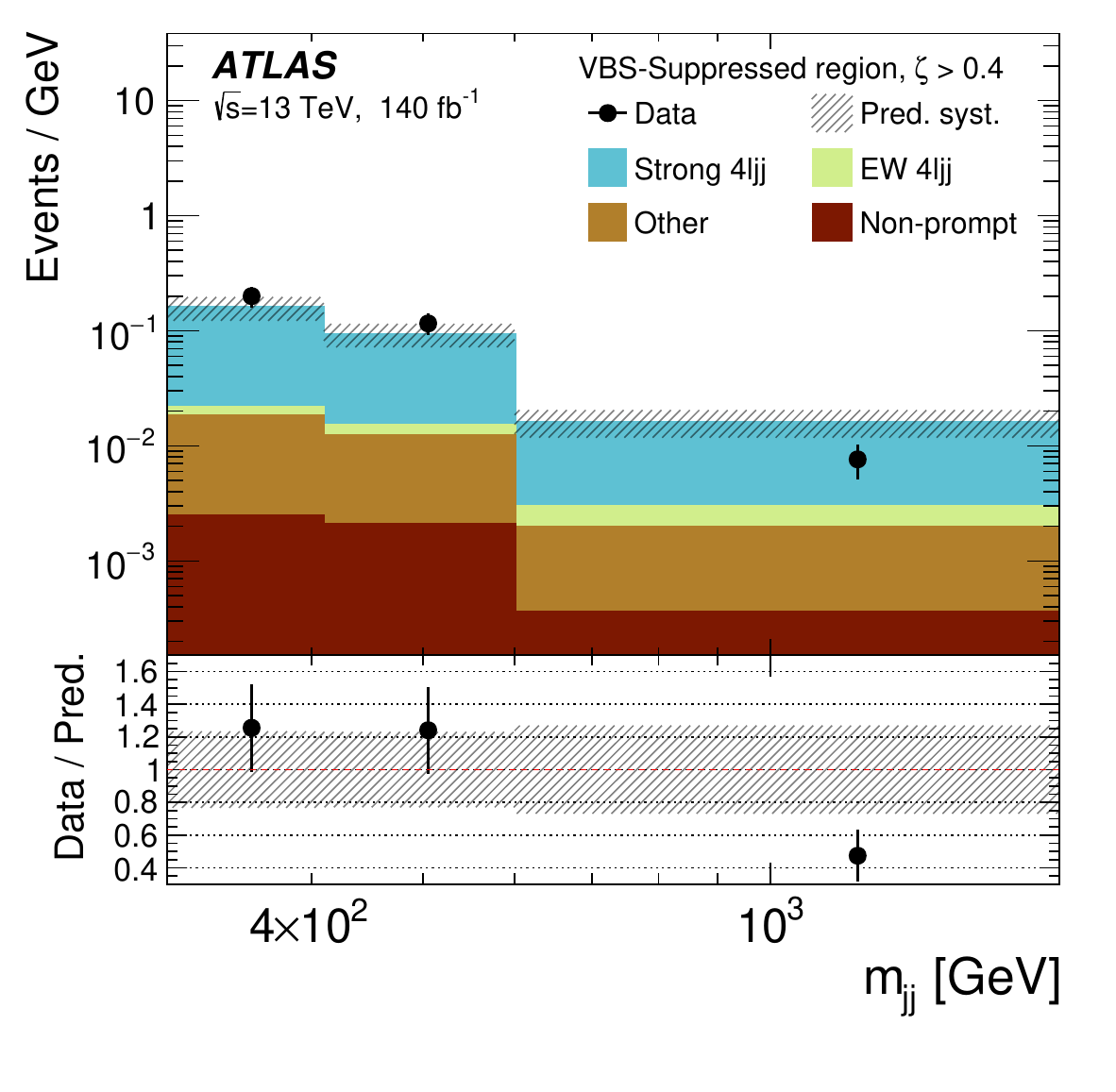}\\
\includegraphics[width=0.5\textwidth]{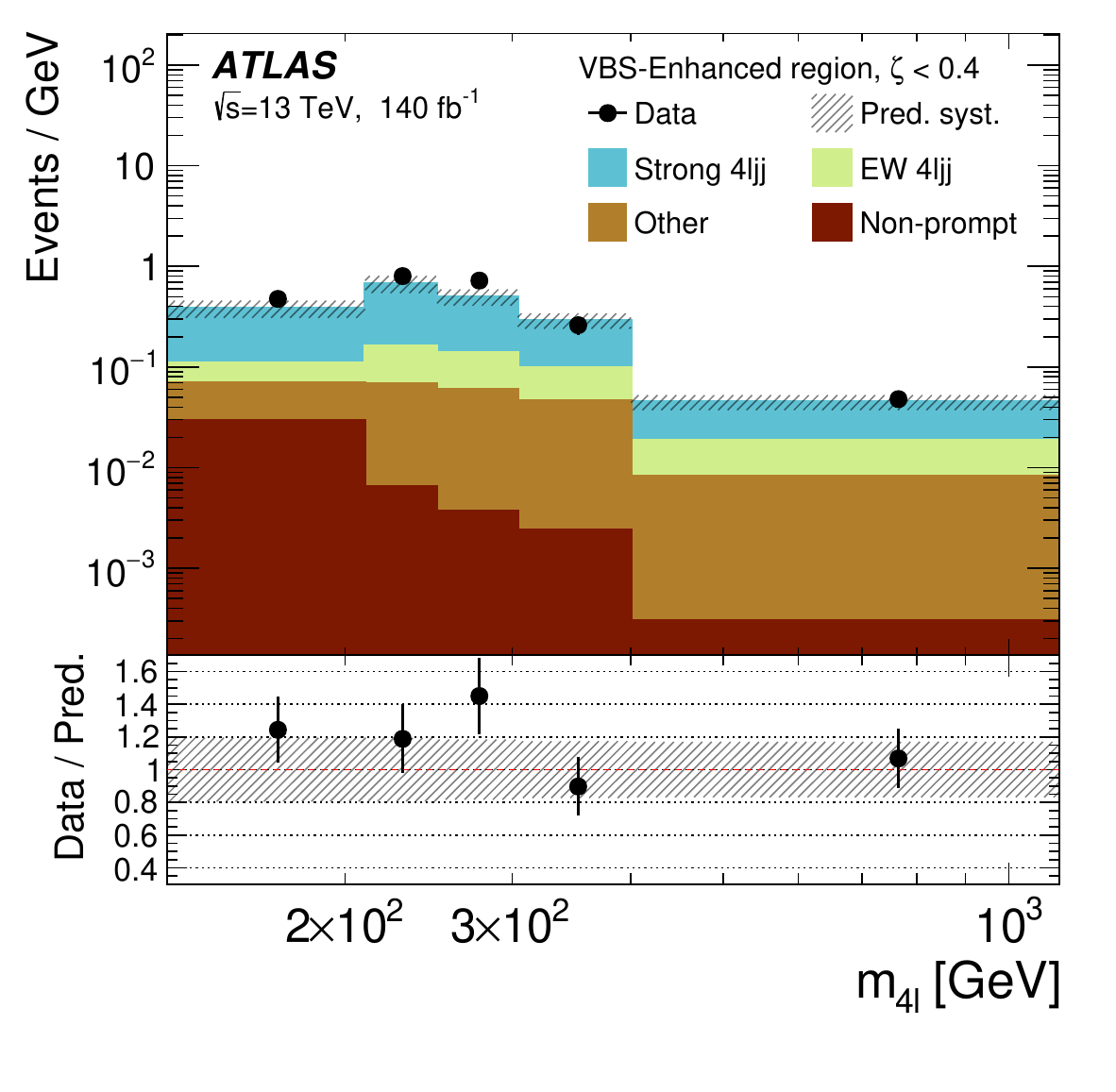}
\includegraphics[width=0.5\textwidth]{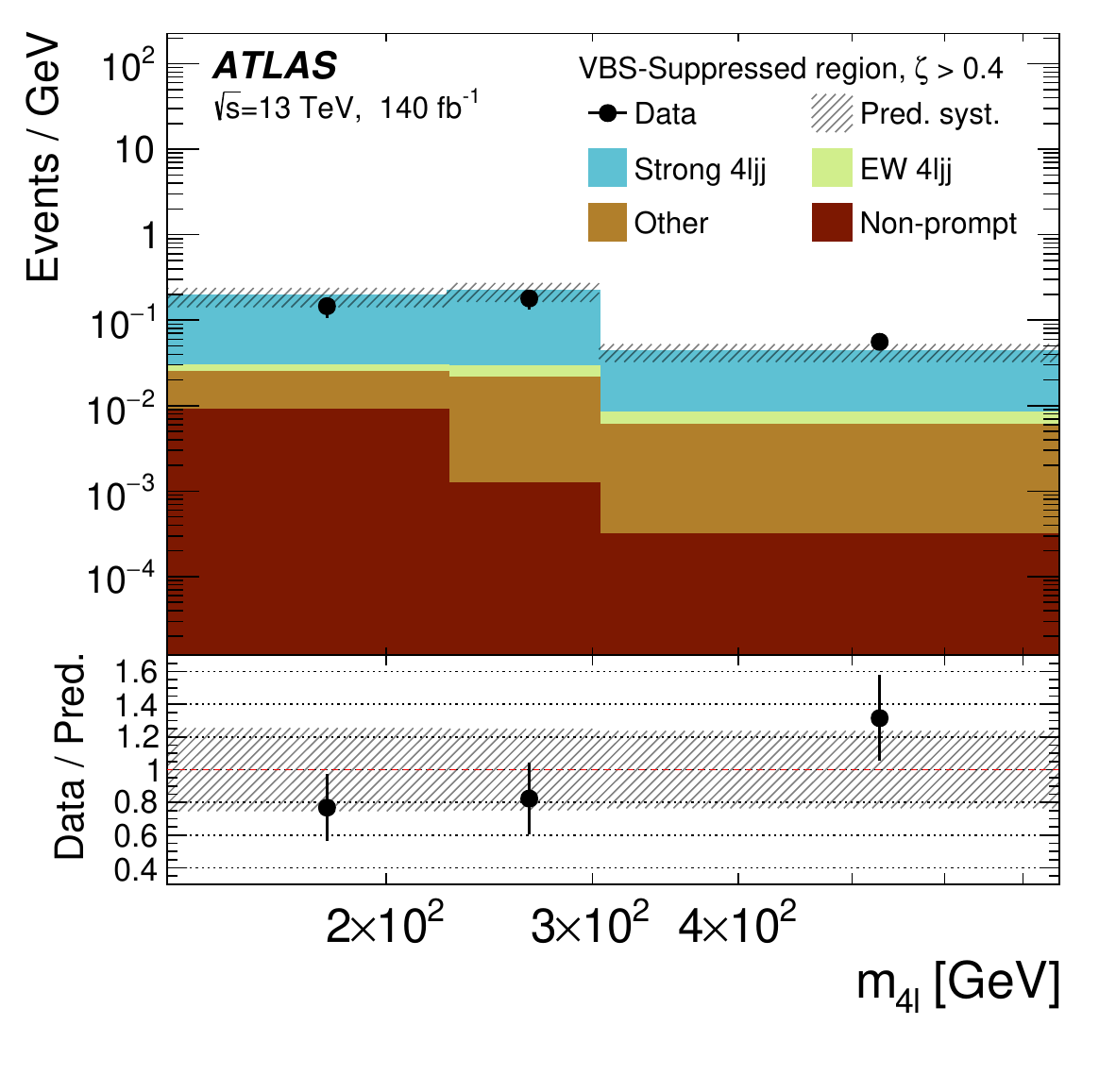}
\caption{Predicted and observed yields as a function of \mjj{} (top) and \mfl{} (bottom), measured in the VBS-enhanced (left) and VBS-suppressed (right) regions. The data are represented as black points and the associated error bars represent the statistical uncertainty.
The background arising from non-prompt leptons is estimated using a data-driven technique as outlined in Section~\ref{sec:background}. Background processes with four prompt leptons, such as $t\bar{t}Z$ production and the fully leptonic decays of $WWZ$ and $WZZ$ production, are estimated by using simulations and labelled as `Other.'
The total uncertainty on the combined signal and background prediction is shown as a grey band (the calculation of these uncertainties is outlined in Section~\ref{sec:systematics}).
\label{fig:dataMCcomparison}}
\end{figure}


\label{sec:evsel}
 
\section{Backgrounds}

Processes with leptons that do not originate from the decay of a $Z$ boson are considered as backgrounds in this analysis. Background processes with at least four prompt leptons, such as $t\bar{t}Z$ production and the fully leptonic decays of $WWZ$ and $WZZ$ production, are estimated by using the MC  simulations presented in Section~\ref{sec:datasimul}.
 
Backgrounds that contain one or more non-prompt leptons arise from $WZjj$ production (where an additional jet produces a non-prompt lepton) and $t\bar{t}$ production (where the $b$-hadrons produced in the top/anti-top decay are the main sources of the non-prompt leptons). These backgrounds are typically poorly modelled in simulation and are estimated using a data-driven method, whereby their event yield is measured in a control region enriched in non-prompt leptons and extrapolated to the signal region using a scaling factor based on the non-prompt lepton efficiency.
 
The control region is defined using the same event selection as outlined in Sec.~\ref{sec:evsel}, but with one (or more) of the four leptons that define the $Z$ boson candidates failing to meet the \textit{signal} lepton definition. The event yields per bin of a measured distribution are then extrapolated to the signal region by applying a weight of $f/(1-f)$ for each non-\textit{signal} lepton in each event, where $f$ is the non-prompt lepton efficiency and defined as the fraction of \textit{baseline} non-prompt leptons that satisfy the \textit{signal}-lepton definition.  The contribution to the event yields in the control region from processes with at least four prompt leptons are subtracted before the extrapolation. That contribution is typically at the 5\%--10\% level, though rises to 20\% at the largest \mfl{}.
 
The non-prompt lepton efficiency is calculated in data using $Z + \textrm{jet}$ events and $t\bar{t}$ events. Candidate $Z + \textrm{jet}$ events are required to have a SFOC lepton pair with an invariant mass within 10~\GeV of the $Z$  boson mass, whereas the $t\bar{t}$ candidate events are required to have an opposite-flavour opposite-charge lepton pair. The events are then required to have one additional baseline lepton. The non-prompt lepton efficiency is calculated using this additional lepton, after correcting the event yields for contributions from processes that produce at least three prompt leptons, as a function of the lepton $p_\textrm{T}$, the lepton $\eta$ and the number of jets in the event. The non-prompt leptons in $Z + \textrm{jet}$ events arise mainly from light-flavour decays, whereas they arise mainly from heavy-flavour decays in $t\bar{t}$ events. The two non-prompt lepton efficiency measurements are therefore combined to reflect the expected flavour composition of non-prompt leptons in the signal region of this analysis, as estimated by using the simulations presented in Section~\ref{sec:datasimul}.


\label{sec:background}
 
\section{Correction for detector effects}
\label{sec:unfolding}

 
Particle-level differential cross-sections for \zzjj{} production in the VBS-enhanced and VBS-suppressed regions are obtained by correcting the background-subtracted event yields for the effects of detector inefficiency and resolution.

The particle-level fiducial phase space is defined using stable final-state particles with a lifetime of $c\tau>10$~mm. Dressed leptons are used to define the $Z$ boson candidates. Electrons and muons are required to be prompt (i.e., to not originate from the decay of a hadron) and the dressed lepton is defined as the four-vector sum of the electron (or muon) and all prompt photons within a range of $\Delta R < 0.1$.  Dressed electrons are required to have $p_{\textrm{T}}>7$~\GeV\@ and $|\eta|<2.47$. Dressed muons are required to have $p_{\textrm{T}}>5$~\GeV\@  and $|\eta|<2.7$. Jets are defined using the anti-$k_t$ algorithm with radius parameter set to 0.4. All stable particles except dressed leptons are used as input to the jet-finding algorithm. Jets are required to have $p_{\textrm{T}}>30$~\GeV and $|\eta|<4.5$.
 
To minimise model-dependence in the final result, the kinematic criteria used to define the particle-level fiducial regions are similar to the criteria applied to the reconstructed data in Sec~\ref{sec:evsel}. At least four dressed leptons are required and the two leptons with the largest transverse momenta are required to have $p_{\textrm{T}}>20$~\GeV. All possible SFOC lepton pairs are formed and later ordered by $|m_{\ell\ell} - m_Z|$. The two $Z$ boson candidates are defined as the two SFOC pairs with the smallest value of $|m_{\ell\ell} - m_Z|$ that are constructed from different leptons. The leading $Z$ boson candidate is defined as the one with the largest value of  $|y_{\ell\ell}|$. The invariant mass of the four leptons is required to satisfy $\mfl>130$~\GeV. Events are required to contain at least two jets. The dijet system is then defined as the two leading (highest transverse momentum) jets in the event that have $\eta_{j_1} \times \eta_{j_2}<0$. The highest transverse momentum jet is required to have $p_{\textrm T}>40$~\GeV{}  and the dijet system is required to satisfy $\Dyjj>2.0$ and $\mjj>300$~\GeV. Measurements are carried out in two fiducial regions, with the VBF-enhanced and VBF-suppressed regions defined by $\zeta<0.4$ and $\zeta>0.4$, respectively, where $\zeta$ is the centrality of the diboson system defined in Equation~\ref{eq:cent}.
 
The event yields are binned to ensure a minimum of 20 (15) events in each bin in the VBF-enhanced (VBS-suppressed) regions at detector level. The bins are also required to be larger than twice the detector resolution, which is determined using the simulated \zzjj{} events. For non-angular observables (e.g. \mjj{}, \mfl{}), the last bin is chosen such that less than one \zzjj{} event is predicted to lie above the upper bin edge. However, the `overflow' events that lie above that bin edge are included in the last bin. The background-subtracted event yields are then corrected to particle level using an iterative Bayesian unfolding method~\cite{DAGOSTINI1995487,Adye:2011gm}. The measurements are unfolded using two iterations, except for the \pTfljj{} distribution in the VBF-enhanced region for which three iterations are used.\footnote{The number of iterations is optimised so the quadrature sum of the statistical uncertainty and the unfolding bias uncertainty (discussed in Sec.~\ref{sec:systematics}) is minimised, whilst also requiring that the per-bin statistical uncertainty after unfolding is not smaller than that observed at detector level.} The unfolding method uses simulations to (i) correct for events selected at detector level that do not satisfy the particle-level selection, (ii) correct for migrations between bins of the measured spectrum with a response matrix, and (iii) correct for events selected at particle level but not at detector level.
 
The default simulations used in the unfolding are the EW \zzjj{} sample produced using \mglo{} and the strong \zzjj{}  samples produced using \textsc{Sherpa}. The fraction of detector-level events that are also reconstructed at particle level, referred as the fiducial fraction, is typically between 60\% and 80\%. About half of the non-fiducial events have at least one pile-up jet reconstructed in the dijet system, and the remaining non-fiducial events arise from low-$p_\textrm{T}$ jets migrating into the fiducial volume due to jet resolution effects. The fraction of particle-level events that are also reconstructed at detector level, referred to as the efficiency, is typically between 40\% and 60\%. The low overall efficiency is due to inefficiencies in the trigger and the lepton reconstruction plus identification and isolation criteria. The migration among bins is typically small due to the binning methodology, and ranges from 10\% to 30\%.
 
Statistical uncertainties in the data are propagated through the unfolding using a bootstrap method~\cite{PhysRevD.39.274}, with 10 000 pseudo-experiments. For each pseudo-experiment, each event is randomly assigned a weight based on a Poisson distribution with a mean of one.  The unfolding is then repeated using the modified distributions created from the event weights. The final statistical uncertainties in the measurement are taken to be the standard deviation of the unfolded values obtained from the ensemble of pseudo-experiments. The statistical uncertainties in the \zzjj{} simulations are propagated through the unfolding procedure in a similar fashion, whereby each pseudo-experiment modifies the simulations used in the unfolding procedure.
 
To ensure that the unfolding procedure is unaffected by the presence of beyond-the-SM physics in the data, a signal injection test has been performed using the effective field theory (EFT) predictions presented in Section~\ref{sec:eft}. Pseudo-data from the EFT simulations are unfolded with the response matrices constructed from the nominal SM simulation, and the result is compared with the particle-level EFT simulation. The unfolding procedure is shown to introduce very little bias, with differences between the unfolded and particle-level distributions being much smaller than the experimental uncertainties in the measurement.


\section{Systematic uncertainties}
\label{sec:syst}


\textbf{Experimental systematic uncertainties} arise from differences between data and simulation in the trigger efficiency, the lepton reconstruction, the jet reconstruction, the luminosity determination, and the average number of proton--proton interactions per beam bunch crossing. These uncertainties affect the normalisation of the background processes estimated by using simulations and the simulations of the signal processes used in the unfolding procedure. For each source of uncertainty, the simulations are varied by $\pm1$ standard deviations and the analysis repeated. The systematic uncertainty is defined as the change in the differential cross-section relative to the nominal analysis with the default simulations.
 
The luminosity of the data sample is known to an accuracy of 0.83\% using a combination of van der Meer beam separation scans in dedicated running periods and luminosity-sensitive detectors in standard data-taking periods \cite{ATLAS:2022hro}.
 
The average number of proton--proton interactions per bunch crossing varies during the different data-taking periods, and the simulations are corrected to reproduce the distribution observed in the data. This procedure is impacted by an uncertainty in the ratio of the predicted and measured inelastic cross-sections within the ATLAS fiducial volume~\cite{ATLAS:2016ygv}. A systematic uncertainty is introduced to account for this, by scaling the average number of proton--proton interactions per bunch crossing in the simulation.
 
The lepton reconstruction, identification, isolation and trigger efficiencies in simulation are corrected, using scale factors, such that they match those observed in the data, as outlined in Section~\ref{sec:datasimul}. Systematic uncertainties in this procedure are estimated by modifying the scale factors by their associated uncertainties~\cite{MUON-2018-03,EGAM-2018-01}. Uncertainties in the lepton momentum scale and resolution are estimated by scaling and smearing the lepton transverse momentum by the known differences between data and simulation~\cite{ATLAS:2022jjr,EGAM-2018-01}.
 
Jets are calibrated using a combination of MC-based and data-driven corrections, as discussed in Section~\ref{sec:evsel}. The jet energy scale and jet energy resolution uncertainties are estimated by scaling and smearing the jet four-momentum in the simulation by the associated uncertainties in the calibration procedure~\cite{JETM-2018-05}. Furthermore, the JVT algorithm introduces an inefficiency in the jet reconstruction. The uncertainty that arises from imperfect modelling of the JVT in simulation is estimated by varying the JVT requirement~\cite{PERF-2014-03, PERF-2016-06}.
 
\textbf{Backgrounds} containing at least four prompt leptons are normalised to ATLAS measurements, as outlined in Section~\ref{sec:datasimul}. The uncertainty that arises from subtracting these backgrounds is estimated by varying the normalisation of the simulated samples by an amount commensurate with the experimental precision of the ATLAS measurements. Backgrounds containing non-prompt leptons are estimated by using the data-driven method outlined in Section~\ref{sec:background}. The non-prompt background estimate contains statistical uncertainties from the finite event yields in the control regions and systematic uncertainties from the subtraction of prompt-lepton processes from those event yields. The uncertainty associated with subtracting the non-prompt backgrounds is estimated by varying the background estimates by an amount commensurate with the statistical and systematic uncertainty. Uncertainties associated with the expected flavour composition of non-prompt leptons in the signal region have been checked and found negligible.

\textbf{Theoretical systematic uncertainties} in the simulation of \zzjj{} production impact the measurement via the unfolding procedure. Three sources of uncertainty are investigated, arising from (i)  the  renormalisation and factorisation scale dependence in the \zzjj{}  calculations, (ii) the parton distribution functions, and (iii) the choice of event generator.
 
The default simulations used in the unfolding procedure are \mglo{} (EW \zzjj{}) and \textsc{Sherpa} (strong \zzjj{}). These predictions are produced with a default choice for factorisation scale, $\mu_F$, and renormalisation scale, $\mu_R$, but also have additional weights that allow the impact  of renormalisation and scale variations to be assessed. There are six variations in total, corresponding to $\{\mu_R, \mu_F\}$ being scaled by $\{0.5, 0.5\}$, $\{0.5, 1.0\}$, $\{1.0, 0.5\}$, $\{1.0, 2.0\}$, $\{2.0, 1.0\}$, or $\{2.0, 2.0\}$. The systematic uncertainty due to scale choice for each generator prediction is then taken to be the envelope of the six scale variations. However, for the loop-induced $gg \rightarrow 4\ell$ contribution to the strong \zzjj{} process, the normalisation uncertainty is taken from the higher-order calculations to which the MC sample is normalised, and the scale variations are only used to assess the additional uncertainty in the shape of each distribution.

Uncertainties associated with PDFs are evaluated using the PDF4LHC recommendation~\cite{Butterworth:2015oua}. First, the nominal prediction for each generator is reweighted by the 100 variations of the NNPDF PDF set (including the associated \alphas{} variations). The standard deviation of the 100 predictions that are obtained is taken as the uncertainty in the NNPDF prediction. The nominal prediction is then reweighted to reproduce the nominal MMHT and CT14 NNLO PDF sets. The total PDF uncertainty for a given process is then taken as the envelope of the NNPDF uncertainty and the MMHT / CT14 predictions.

The dependence on the choice of event generator is evaluated by unfolding the Asimov data sample constructed from the \mglo{} strong \zzjj{} simulation with the nominal response matrix. The \mglo{} sample is first reweighted on an event-by-event basis to reproduce the \sherpa{} prediction at particle level, which avoids double counting uncertainties associated with the sample normalisation that is obtained from the factorisation and renormalisation scale variations. The difference between the unfolded event yield and the reweighted \mgnlo{} particle-level prediction is taken as a symmetric systematic uncertainty in the differential cross-section measurement.
 
The effect of interference between the EW \zzjj{} and strong \zzjj{} processes is estimated at leading-order in perturbative QCD using \MGNLO~2.6.1. The contribution is found to be far smaller than the theoretical uncertainties in the EW \zzjj{} and strong \zzjj{}  calculations and no additional uncertainty is added.

\begin{table} [tp]
\centering
\begin{tabular}{|l|c|c|}
\hline \hline
Source & \multicolumn{2}{c|}{Uncertainty (\%)}\\
& VBS-enhanced region & VBS-suppressed region \\
\hline
Luminosity & 0.8 -- 2.1 & 0.8 -- 2.0 \\
Leptons  & 0.8 -- 1.6 & 1.0 -- 1.5 \\
Jets  & 2.7 -- 18 & 3.4 -- 13 \\
Pile-up & 0.0 -- 2.5 & 0.0 -- 0.7 \\
Backgrounds & 0.9 -- 9.0 & 1.2 -- 7.0 \\
Theory modelling & 0.6 -- 7.5 & 1.2 -- 8.8 \\
Unfolding method & 0.9 -- 12 & 1.2 -- 12 \\
\hline
Total systematic & 6 -- 22 & 5 -- 17 \\
\hline \hline
\end{tabular}
\vspace{0.5cm}
\caption{Systematic uncertainties in the differential cross-section measurements. The range given for each uncertainty reflects the fact that the uncertainties depend on the underlying kinematics of the process and are therefore larger in some bins of certain distributions. 
}
\label{tab:systsummary}
\end{table}
 
\begin{figure}[tp]
\includegraphics[width=0.5\textwidth]{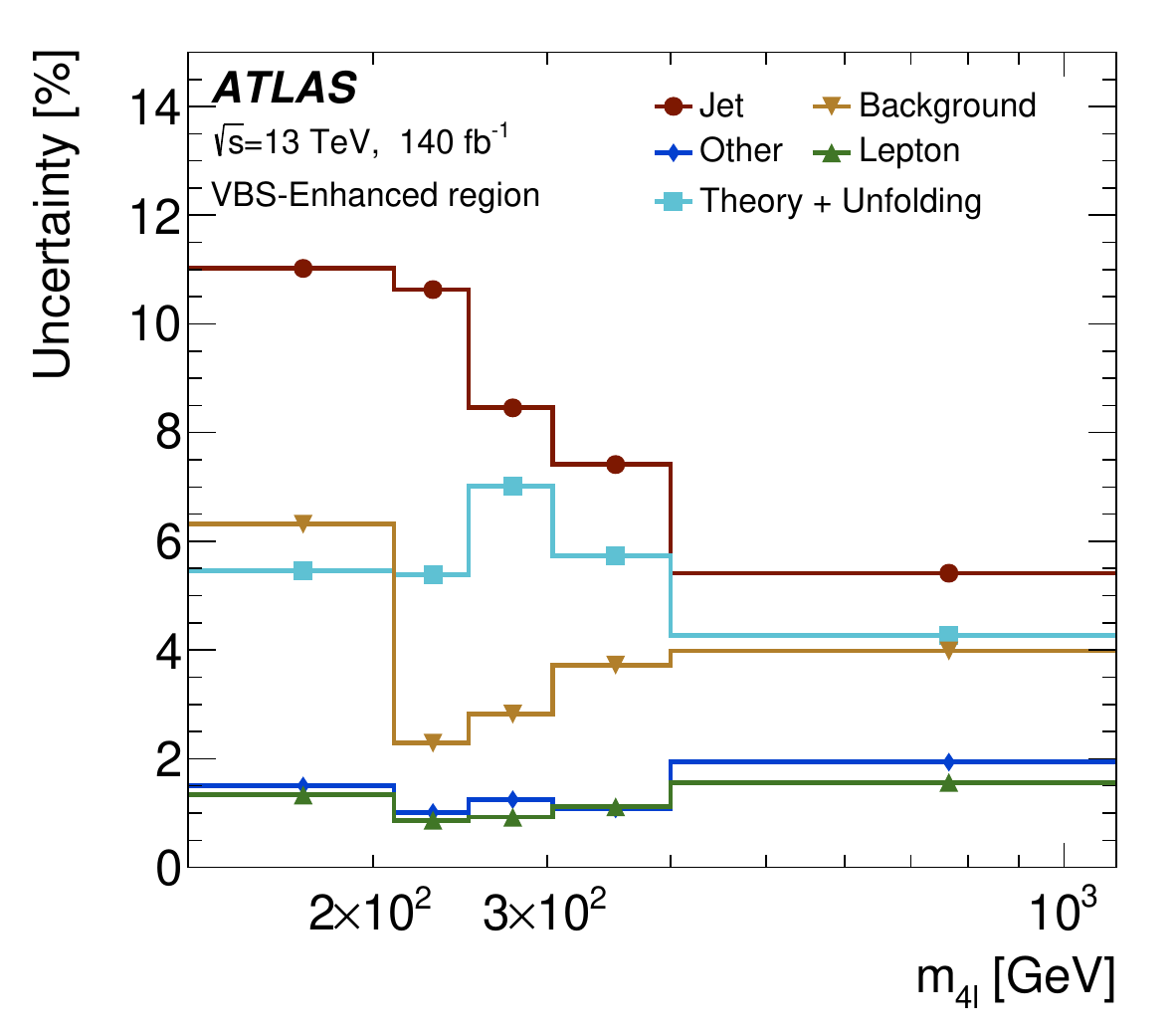}
\includegraphics[width=0.5\textwidth]{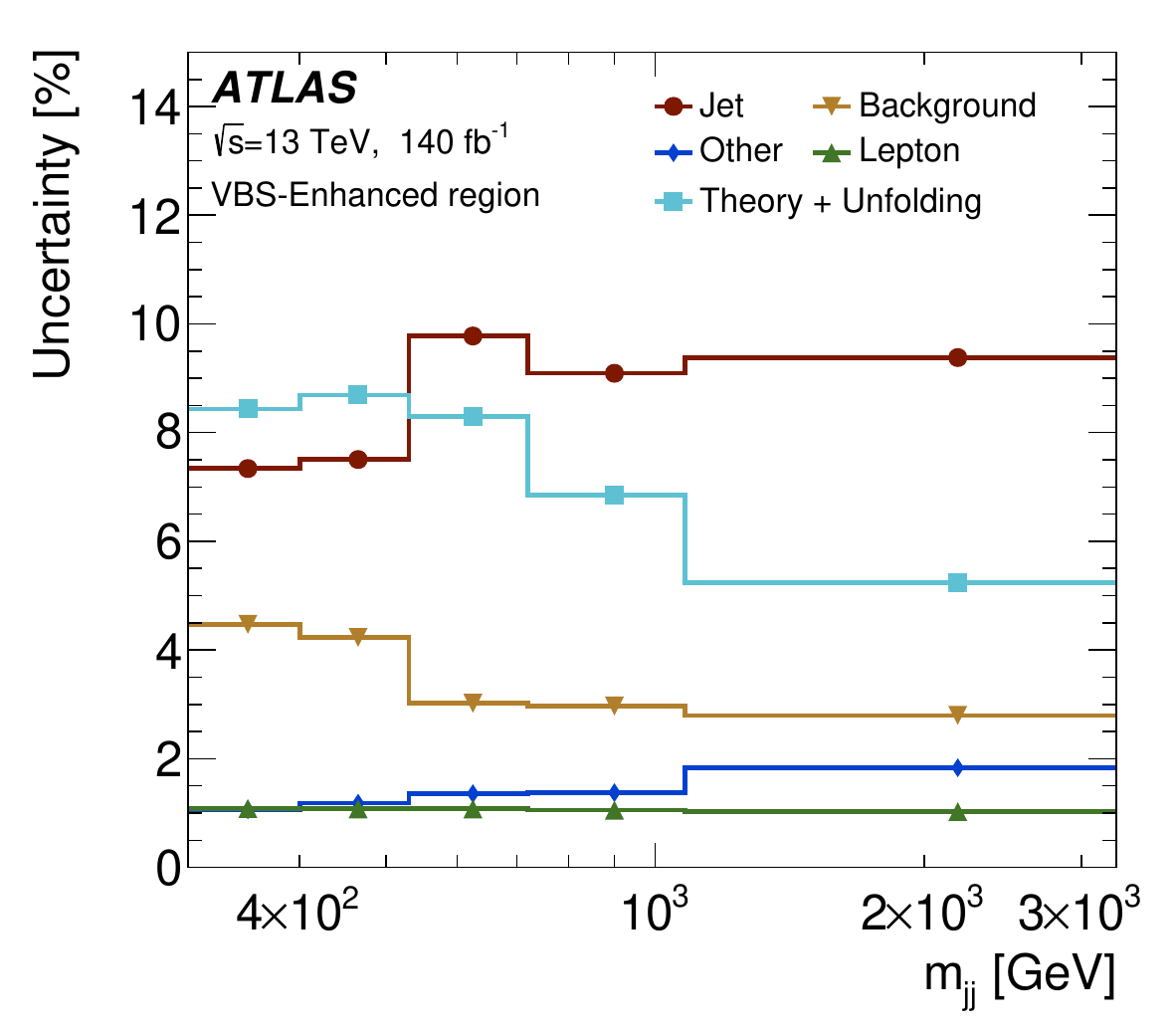}
 
\caption{Systematic uncertainties in the differential cross section for \zzjj{} production in the VBS-enhanced region as a function of \mfl{} (left) and \mjj (right).
\label{fig:syst-breakdown}}
\end{figure}
 
\textbf{Systematic uncertainties in the unfolding method} are evaluated in two ways, using bias tests implemented in simulation. In the first approach, $10\;000$ pseudo-experiments are created by independently fluctuating each bin of the expected particle-level event yield for each observable. Each pseudo-experiment is then folded using the nominal response matrix and the nominal yield of `non-fiducial' events is added (non fiducial events are those that satisfy the detector-level selection but not the particle-level selection). Each pseudo-experiment is then unfolded using the nominal response matrix and the bias for that pseudo-experiment is taken to be the difference between the unfolded yield and the particle-level yield for that pseudo-experiment. 
The bias in the unfolding method is taken to be the bias estimated when the particle-level yield has varied by $\pm1$ standard deviation. In the second approach, the particle-level events are reweighted such that the simulation better matches the data at detector level. The reweighted detector-level distribution is then unfolded using the nominal response matrix. The bias in the unfolding is taken to be the difference between the unfolded spectrum and the reweighted particle-level prediction. The two approaches for assessing bias in the unfolding method give similar results, and the uncertainty in the unfolding method is defined using the first approach.

A summary of the experimental and theoretical uncertainties in the differential cross-section measurements is shown in Table~\ref{tab:systsummary}. The kinematic dependence of these uncertainties as a function of \mfl{} and \mjj{} are shown in Figure~\ref{fig:syst-breakdown}. The dominant systematic uncertainties are those associated with the jet reconstruction, the theoretical modelling, and the unfolding method.


\label{sec:systematics}
 
\section{Results}
\label{sec:results}

The differential cross-sections for inclusive \zzjj{} production in the VBS-enhanced region as a function of \mfl{}, \mjj,
\Dphijj{}, \cosTS{}, \pTfljj{}, and \STfljj{} are shown in Figures \ref{fig:unfolded-vbsenh-masses}, \ref{fig:unfolded-vbsenh-angles}, and \ref{fig:unfolded-vbsenh-qcdvar}. The differential cross-sections as a function of \mfl{} and \mjj{} are examples of observables that are typically used to study vector-boson scattering processes. The differential cross-sections as a function of \cosTS{} are sensitive to the polarisation of the leading $Z$ boson candidate, whereas the differential cross-section as a function of \Dphijj{} is sensitive to the charge-conjugation and parity structure of the $WWZ$ and $WWZZ$ interactions.
The differential cross-sections as a function of \pTfljj{} and \STfljj{} are sensitive to the higher-order real emission of quarks and gluons from the \zzjj{} processes.
Differential cross-sections as a function of other kinematic variables (or in the VBS-suppressed region) are presented in the Appendix.
 
The data are compared with two theoretical predictions, constructed from \mglo{} for the EW \zzjj{}  process and either \sherpa{} or \mgnlo{} for the strong \zzjj{}  process. Scale uncertainties in each \zzjj{}  prediction are estimated by varying the renormalisation and factorisation scales used in the matrix-element calculation independently by a factor of 0.5 or 2.0.
Uncertainties in the \zzjj{}  predictions due to PDFs are estimated for \sherpa{} by reweighting the nominal sample to reproduce the 100 variations of the NNPDF PDF sets (including the associated \alphas{} variations) and taking the RMS of these variations, and for \mglo{} following the PDF4LHC recommendations. The impact of PDF-related uncertainties are found to be much smaller than the impact of scale uncertainties.
 
\begin{figure}[tp]
\includegraphics[width=0.5\textwidth]{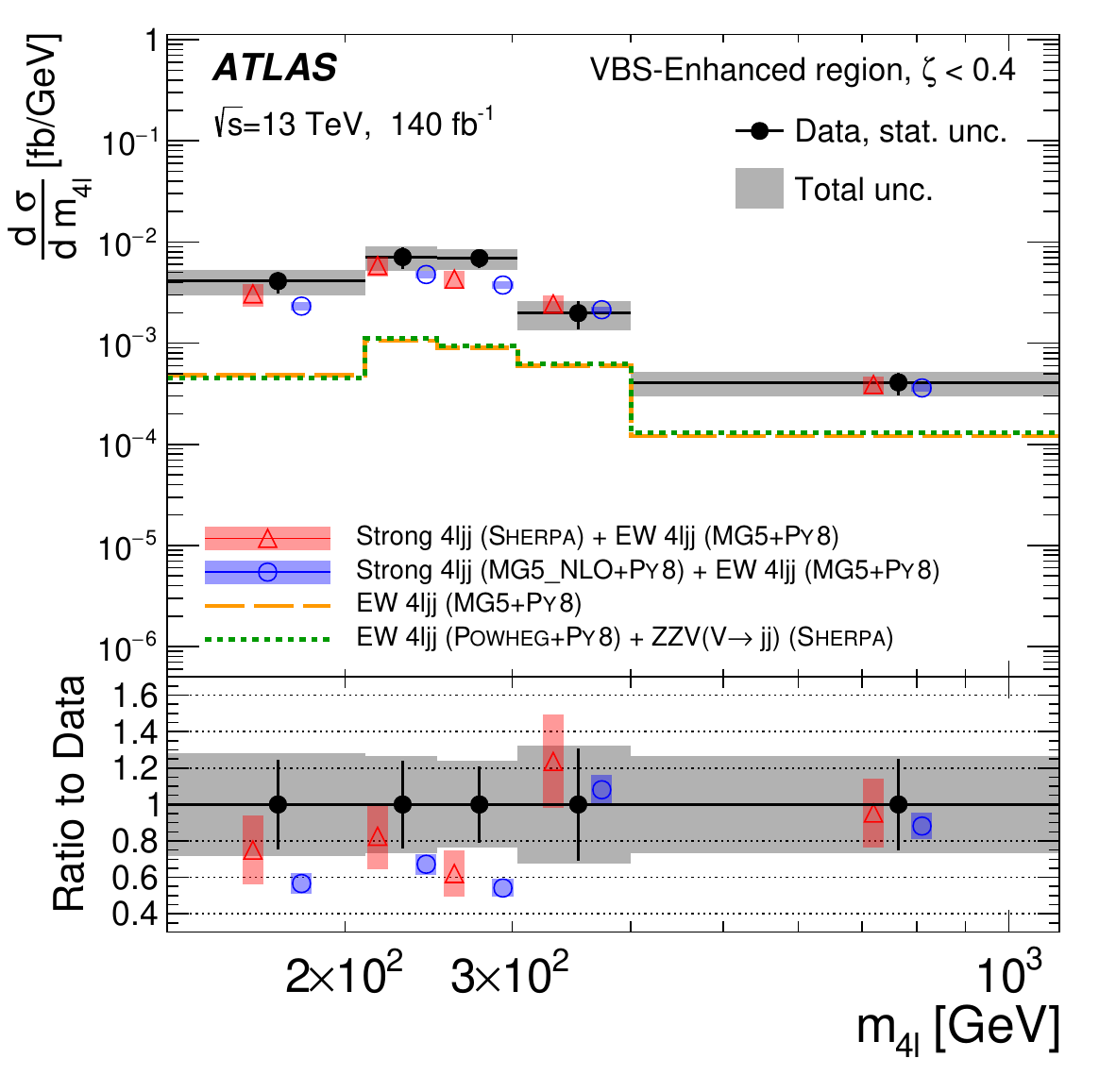}
\includegraphics[width=0.5\textwidth]{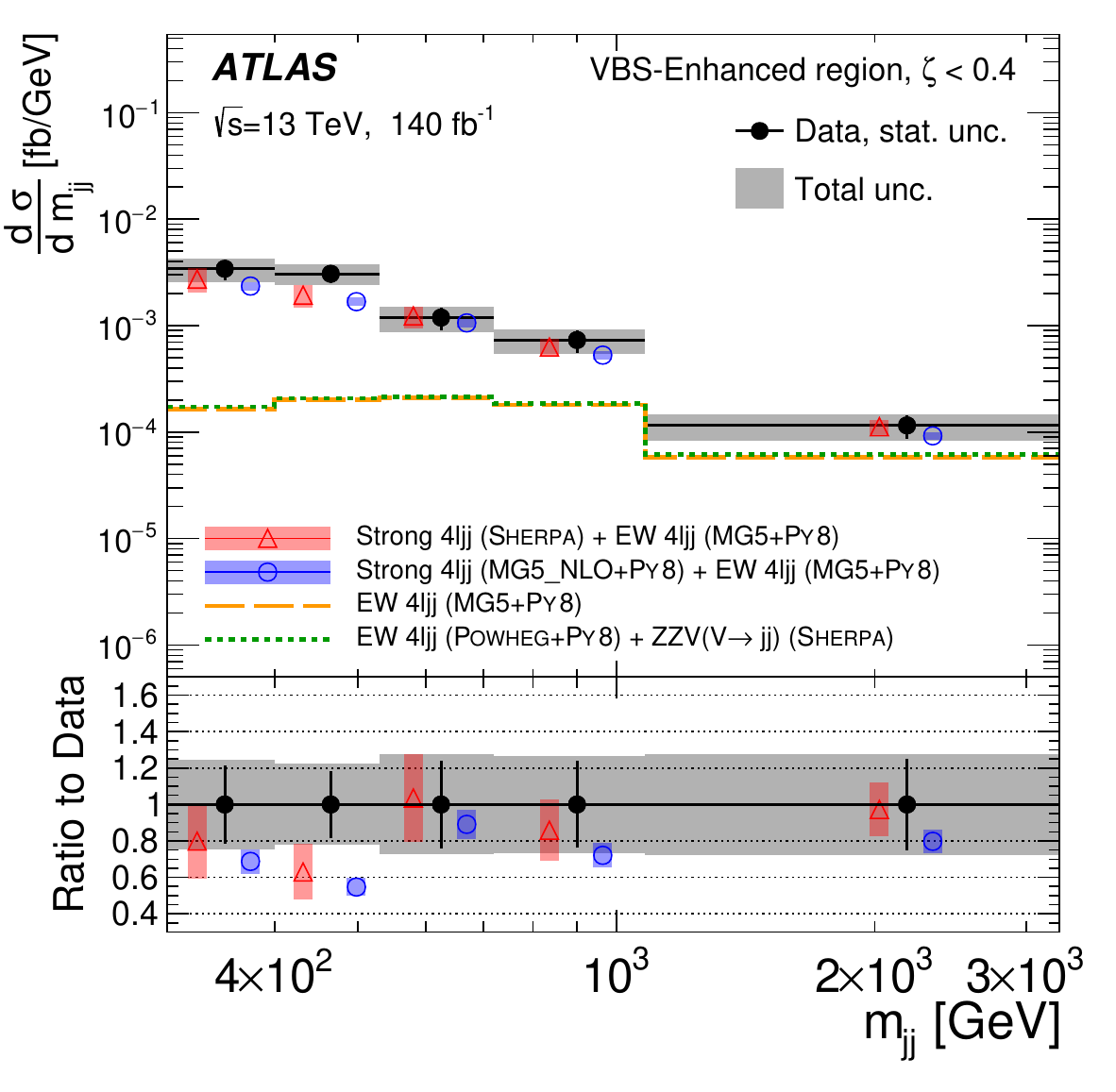}
 
\caption{Differential cross-sections for inclusive \zzjj{}  production in the VBS-enhanced region as a function of
\mfl{} (left) and \mjj (right).
The data are represented as black points and the associated error bars represent the statistical uncertainty. The total uncertainty in the measurement is represented as a grey hatched band. The data are compared with two theoretical predictions, estimated by using \sherpa{} (triangles) and \mg{} (circles) for the strong \zzjj{} contribution and \mglo{} for the EW \zzjj{}  contribution. The band on the theoretical predictions represents the theoretical uncertainty from renormalisation/factorisation scale choices and PDF choice. The dashed lines show the contribution of EW \zzjj{} production to the differential cross-section as predicted by \mglo{} and \powpyt{}. The $s$-channel contributions from $ZZV$ production are missing from the \powpyt{} prediction and are estimated with
\sherpa{}. `Overflow' events that lie above the upper bin edge of the last bin are included in that bin.
\label{fig:unfolded-vbsenh-masses}}
\end{figure}
 
\begin{figure}[tp]
\includegraphics[width=0.5\textwidth]{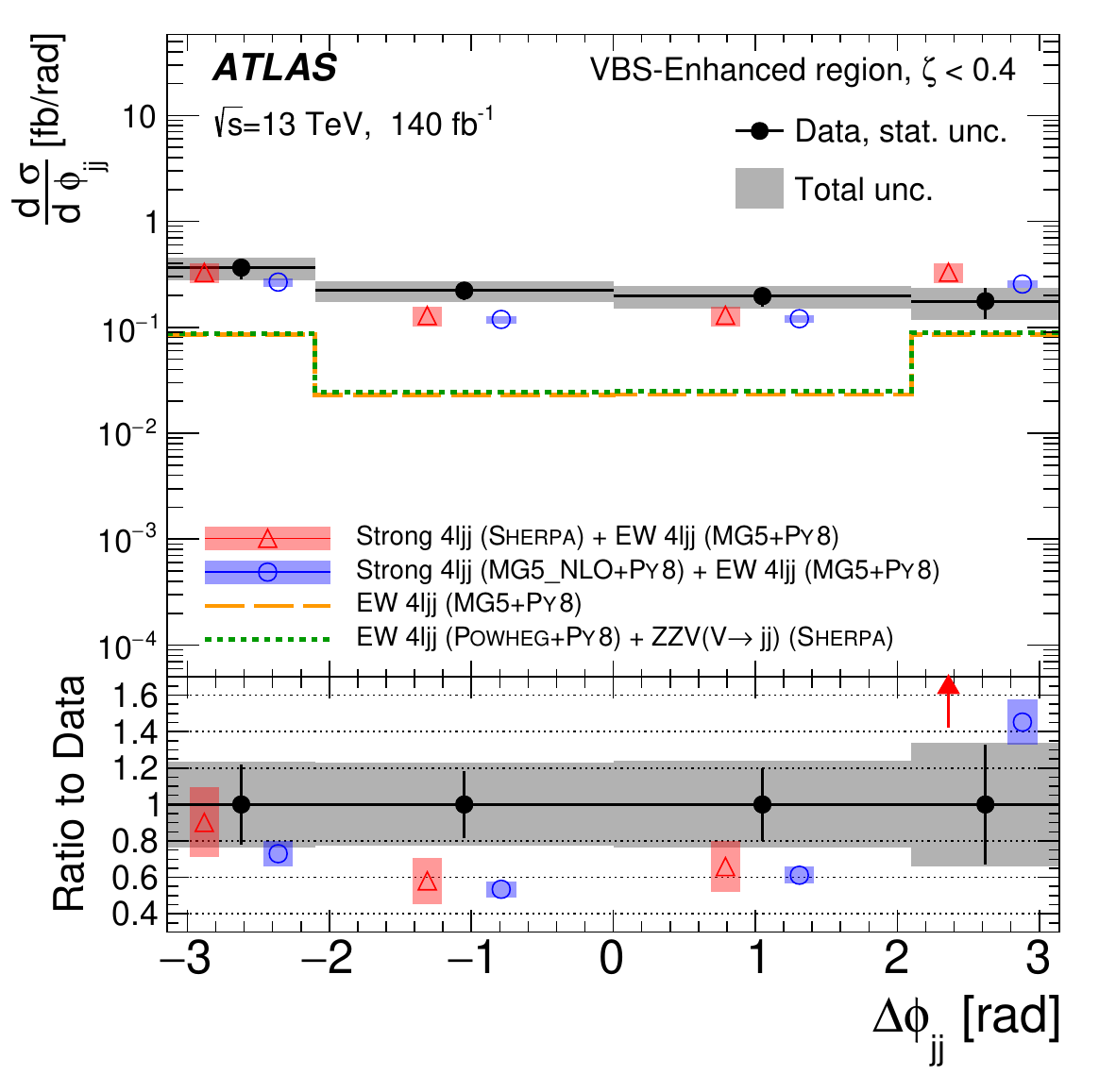}
\includegraphics[width=0.5\textwidth]{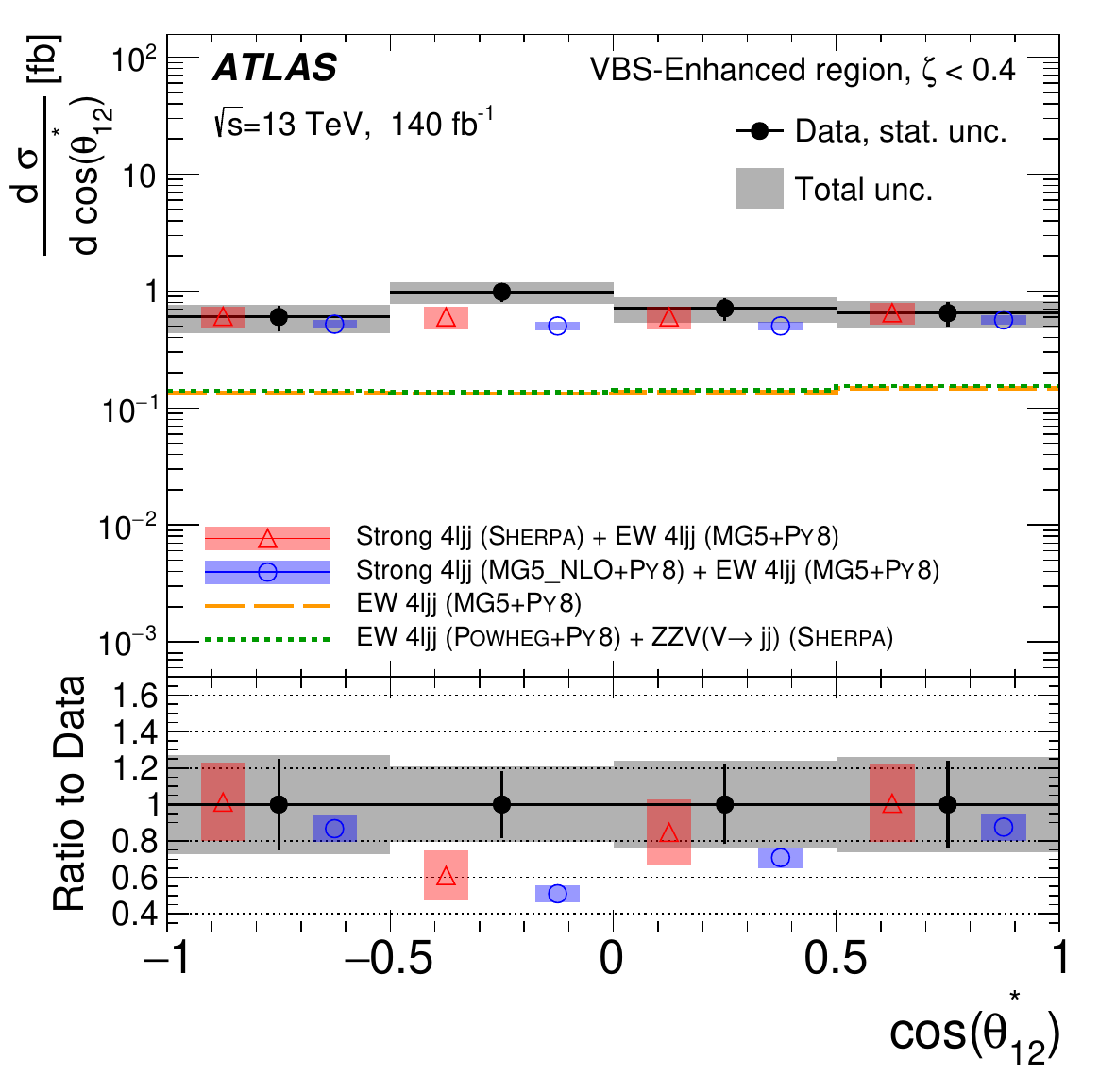}
 
\caption{Differential cross-sections for inclusive \zzjj{}  production in the VBS-enhanced region as a function of
\Dphijj{} (left) and \cosTS{} (right).
The data are represented as black points and the associated error bars represent the statistical uncertainty. The total uncertainty in the measurement is represented as a grey hatched band. The theoretical predictions are constructed in the same way as in Figure~\ref{fig:unfolded-vbsenh-masses}.
\label{fig:unfolded-vbsenh-angles}}
\end{figure}
 
\begin{figure}[tp]
\includegraphics[width=0.5\textwidth]{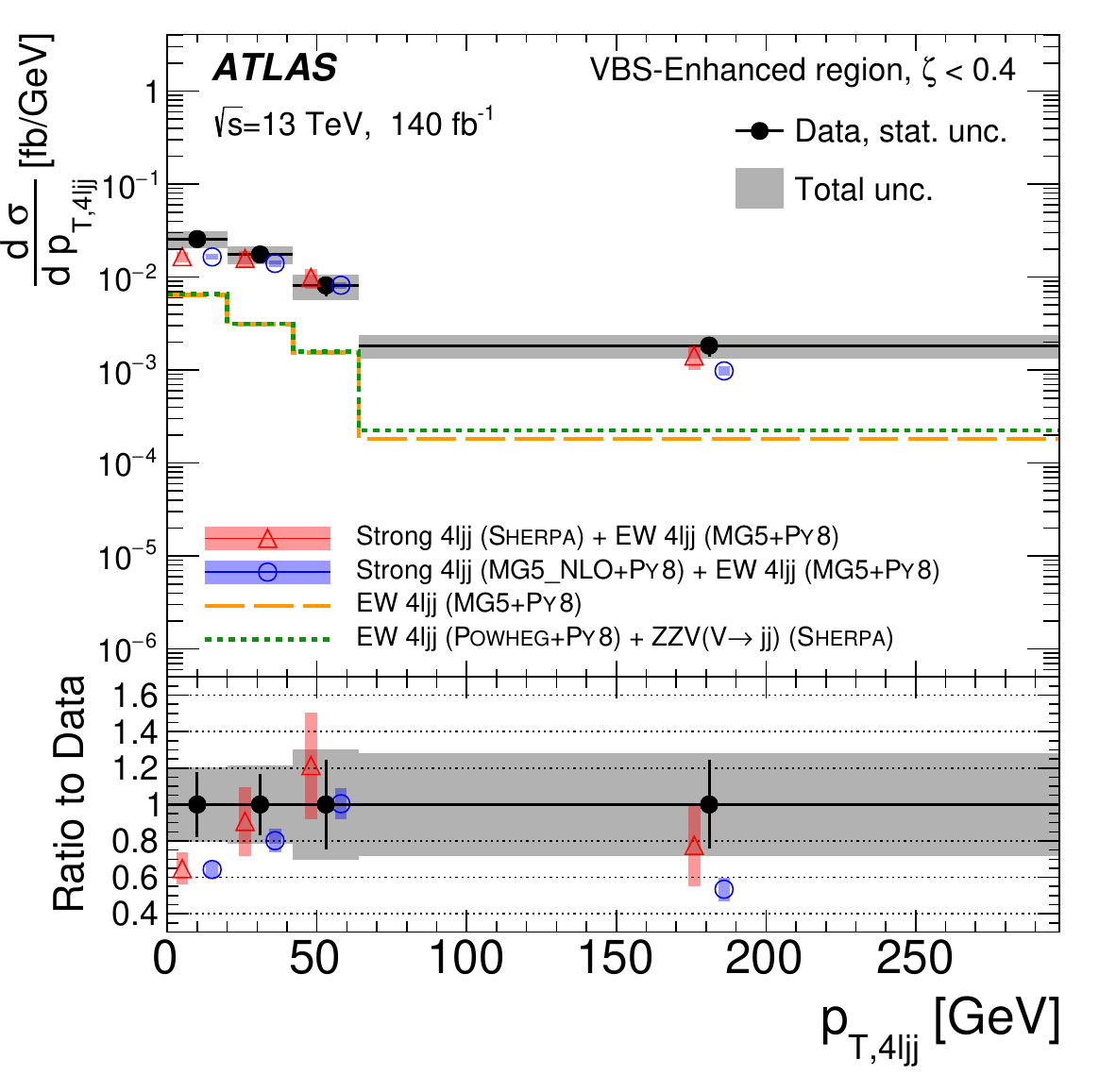}
\includegraphics[width=0.5\textwidth]{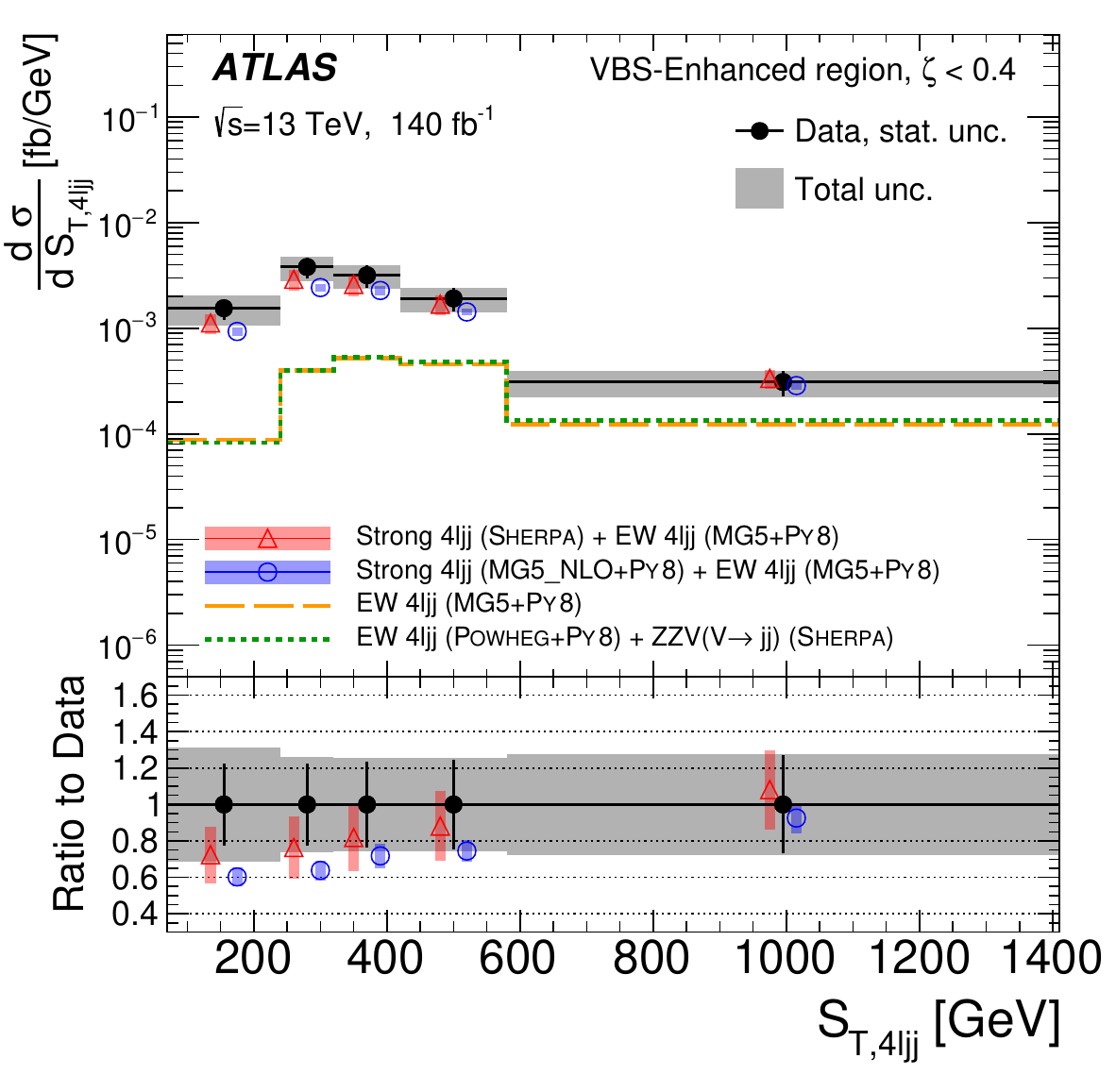}
 
\caption{Differential cross-sections for inclusive \zzjj{}  production in the VBS-enhanced region as a function of
\pTfljj{} (left) and \STfljj{} (right).
The data are represented as black points and the associated error bars represent the statistical uncertainty. The total uncertainty in the measurement is represented as a grey hatched band. The theoretical predictions are constructed in the same way as in Figure~\ref{fig:unfolded-vbsenh-masses}. `Overflow' events that lie above the upper bin edge of the last bin are included in that bin.
\label{fig:unfolded-vbsenh-qcdvar}}
\end{figure}
 
The prediction obtained using \sherpa{} for strong \zzjj{}  production is found to be in satisfactory agreement with the data for all measured distributions in the VBS-enhanced region. However, the prediction obtained using \mglo{} for strong \zzjj{} production is found to underestimate the inclusive \zzjj{} cross-section in all distributions, but the disagreement is especially noticeable at low \mjj{}, low \mfl{} and low $|\Dphijj{}|$. The central values of the \sherpa{} prediction are also below the data, but the agreement with data is improved for \sherpa{} due to the larger theoretical uncertainty in the prediction as estimated from scale variations. The formal accuracy of the calculations is similar, but not identical. Both the predictions are accurate at LO in QCD for the \zzjj{} final state. However, \sherpa{} simulates additional jet activity at LO in QCD whereas \mglo{} relies on the \pythia{} parton shower, which is only accurate to leading logarithm. This may have an impact because the event selection identifies the two `tagging' jets as the highest transverse momentum jets that satisfy $\eta_{j_1} \times \eta_{j_2} < 0$, which may select the third-highest transverse momentum jet in some instances.

The EW \zzjj{}  contribution is also shown separately in Figures~\ref{fig:unfolded-vbsenh-masses}--\ref{fig:unfolded-vbsenh-qcdvar} to show the sensitivity of the measurements to the electroweak process. The electroweak contribution is about 20\% of the measured \zzjj{}  cross-section in the VBS-enhanced region. However, it is much larger at high \mjj{}, where it accounts for about 50\% of the measured cross-section, and much lower at low \mjj{}, where it accounts for just 5\%. Furthermore, the \powpyt{} EW \zzjj{} prediction is in very good agreement with the \mglo{} EW \zzjj{} prediction for all measured distributions, demonstrating that the choice of EW \zzjj{} model has very little impact on the inclusive \zzjj{}  prediction.
 
The differential cross-sections for inclusive \zzjj{} production in the VBS-suppressed region as a function of \mfl{} and \mjj{} are shown in Figure \ref{fig:unfolded-vbssupp}. The data are again compared with the two theoretical predictions that are estimated by using \mglo{} for the EW \zzjj{} process and either \sherpa{} or \mg{} for the strong \zzjj{} process. The electroweak contribution is less than 5\% of the measured \zzjj{} cross-section in this region, and remains below 15\% even at the highest values of \mjj{}. The prediction obtained using \sherpa{} for strong \zzjj{}  production is found to be in satisfactory agreement with the data for the measured distributions, except at the largest values of \mjj{} where the predicted cross section is too large. This feature has been observed in previous measurements sensitive to vector-boson fusion and vector-boson scattering~\cite{ATLAS:2020nzk}. The prediction obtained with \mglo{} is in good agreement with the data at low-\mfl{}, but underestimates the inclusive \zzjj{} cross-section at high-\mfl{} and low-\mjj{}.

\begin{figure}[tp]
\includegraphics[width=0.5\textwidth]{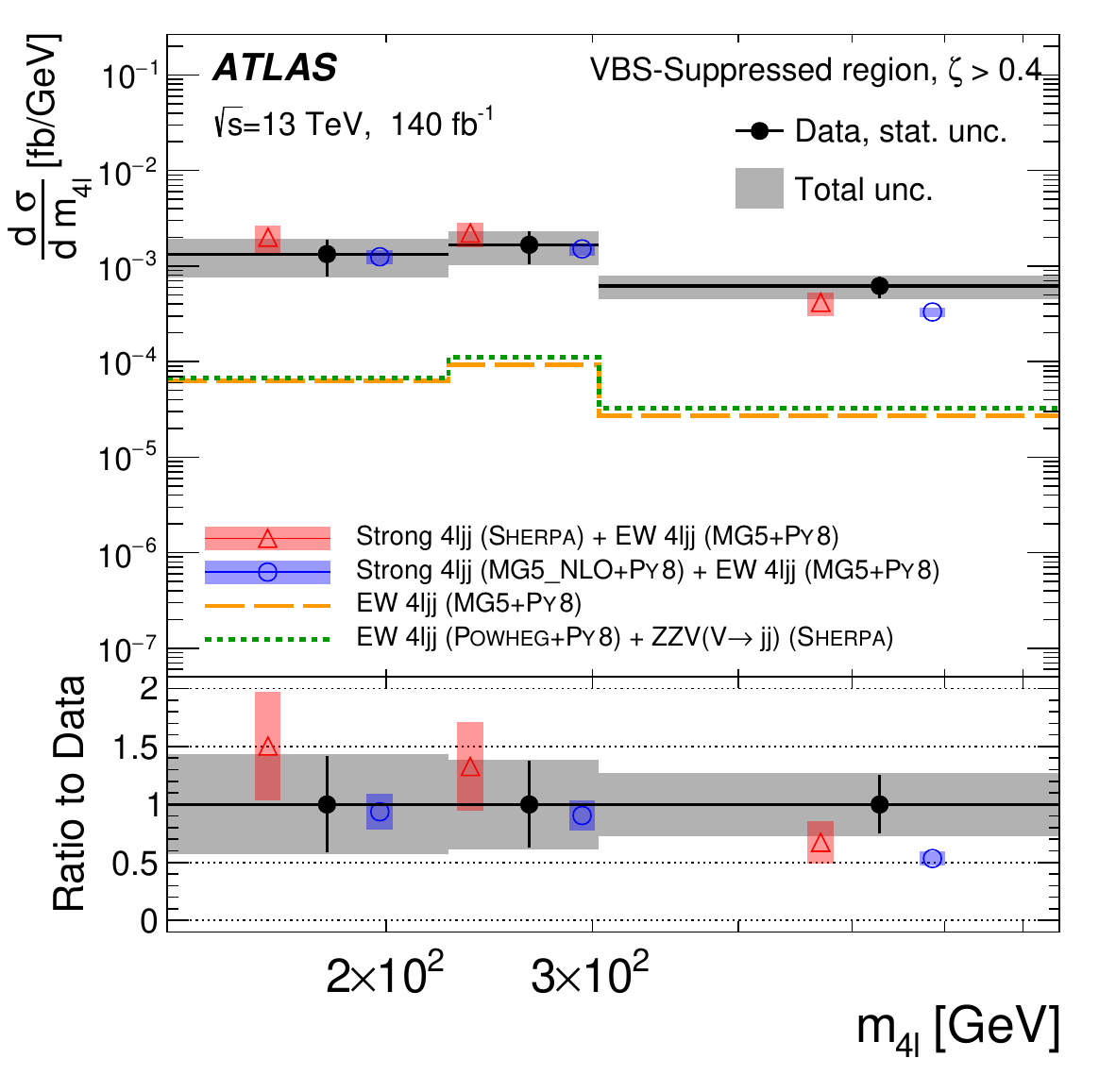}
\includegraphics[width=0.5\textwidth]{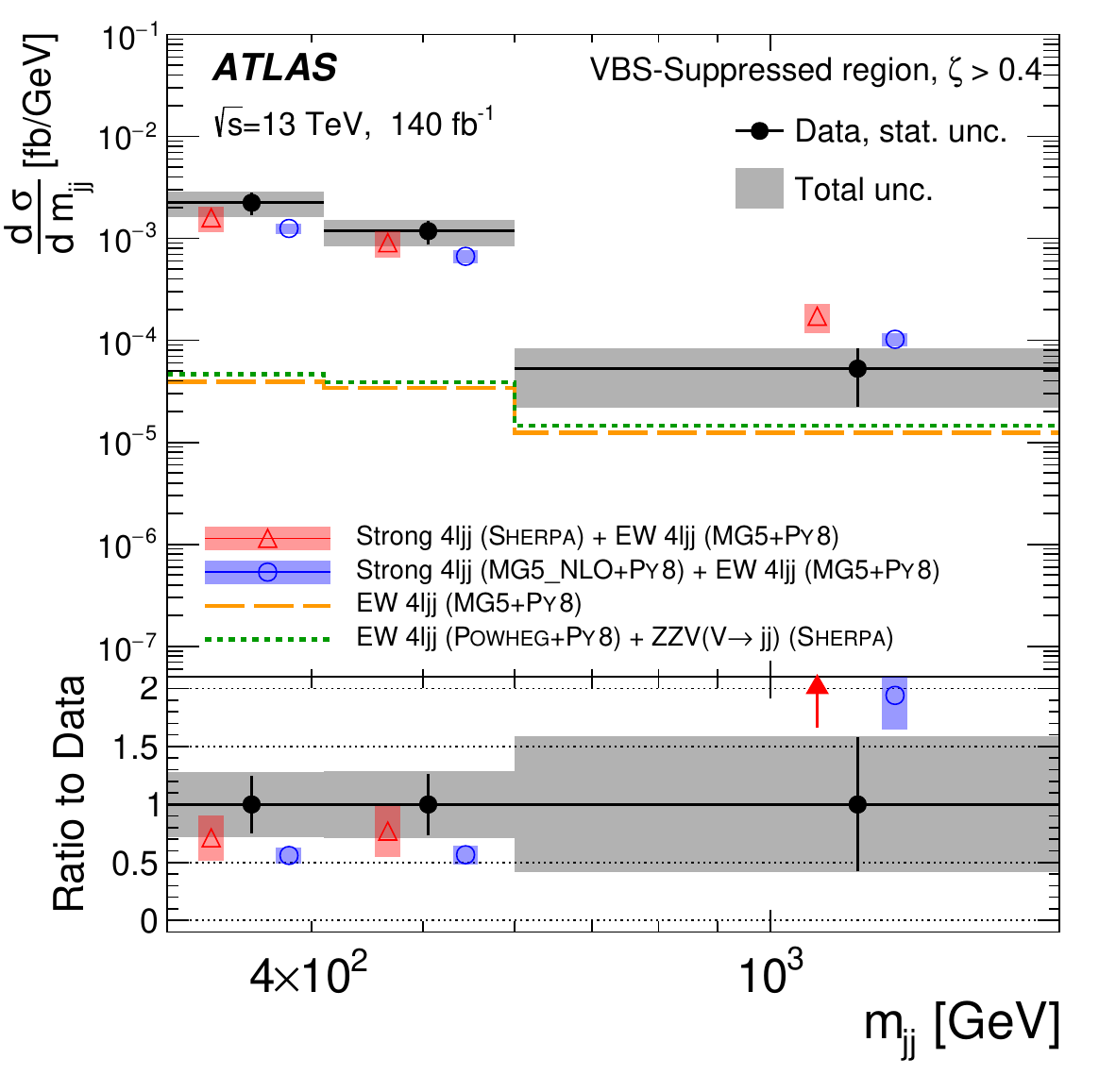}\\
 
\caption{Differential cross-sections for inclusive \zzjj{} production in the VBS-suppressed region as a function of \mfl{} (left) and \mjj (right). The data are represented as black points and the associated error bars represent the statistical uncertainty. The total uncertainty in the measurement is represented as a grey hatched band.
The theoretical predictions are constructed in the same way as for Figure~\ref{fig:unfolded-vbsenh-masses}. `Overflow' events that lie above the upper bin edge of the last bin are included in that bin.
\label{fig:unfolded-vbssupp}}
\end{figure}


\section{Effective field theory interpretation}
\label{sec:eft}

 
The differential cross-sections can be used to search for signatures of physics beyond the SM\@. For measurements sensitive to vector-boson scattering, dimension-eight effective field theory (EFT) modeling can be a tool, whereby the SM Lagrangian is extended with new interactions encoded in dimension-eight operators, i.e.,
\begin{equation*}
\mathcal{L} \, = \, \mathcal{L}_{\textrm{SM}} \, + \, \sum_i \frac{f_{\textrm{T},i}}{\Lambda^4}\mathcal{O}_{\textrm{T},i}
\end{equation*}
where $\mathcal{L}_{\textrm{SM}}$ is the SM Lagrangian, $\mathcal{O}_{\textrm{T},i}$ are a set of the dimension-eight operators, and the $f_{\textrm{T},i}/\Lambda^4$ are Wilson coefficients that specify the strength of the anomalous interactions. The $\mathcal{O}_{\textrm{T},i}$ operators are particularly interesting as they only induce anomalous quartic weak-boson self-interactions~\cite{Eboli:2016kko} and can only be tested using vector-boson scattering processes or triboson production. In focussing on the dimension-eight operators, it is implicitly assumed that the contribution from dimension-six operators is zero, i.e., that they are already constrained from measurements of diboson production~\cite{ATLAS:2019rob,ATLAS:2021kog,CMS:2020mxy, CMS:2021icx, CMS:2021cxr} and vector-boson fusion~\cite{ATLAS:2020nzk}. However, constraints on the Wilson coefficients of operators in a dimension-six effective field theory are presented in the Appendix for completeness.
 
The squared scattering amplitude for the effective field theory prediction can be written as
\begin{equation*}
|\mathcal{M}|^2 = |\mathcal{M_{\textrm{SM}}}|^2 \, + \, 2 \,\textrm{Re} (\mathcal{M_{\textrm{SM}}}^*\mathcal{M_{\textrm{d8}}}) \, + \, |\mathcal{M_{\textrm{d8}}}|^2 ,
\end{equation*}
where $\mathcal{M_{\textrm{SM}}}$ is the SM scattering amplitude and $\mathcal{M_{\textrm{d8}}}$ is a non-SM scattering amplitude that contains anomalous interactions. The cross-section therefore contains three contributions: the SM contribution, the interference between the SM amplitude and the dimension-eight amplitude, and a pure dimension-8 contribution. The interference term and the pure dimension-eight term  contribute at order $f_{\textrm{T},i}/\Lambda^4$ and $f_{\textrm{T},i}^2/\Lambda^8$, respectively. At large values of the Wilson coefficients, the contribution of the pure dimension-eight term can  therefore be larger than the interference contribution, and the theoretical prediction is then sensitive to missing higher-orders in the EFT expansion that contribute at order $1/\Lambda^8$.

Theoretical predictions for the interference  and pure dimension-eight contributions to EW $ZZjj$ production are produced using \mg{}.
Simulated events are produced for each operator separately with the associated Wilson coefficent set to unity. The events are passed through \pythia{} to produce the $ZZ\rightarrow \ell^+\ell^-\ell^+\ell^-$ decay channel and to simulate the fully hadronic final state using the A14 set of tuned parameters. For each operator, the  effective field theory prediction is given by the sum of the interference contribution, the pure-dimension-8 contribution, the \mglo{} EW \zzjj{}  prediction, and the \sherpa{} strong \zzjj{}  prediction. The interference contribution has a linear dependence on the Wilson coefficient, whereas the pure-dimension-8 contribution has a quadratic dependence. This enables modelling of theoretical predictions for any value of a given Wilson coefficient.
 
The measured differential cross-section as a function of \mfl{} and \mjj{} in the VBS-enhanced region and the associated EFT-dependent theoretical predictions are used to define a likelihood function, assuming Gaussian-distributed uncertainties. Statistical correlations between the bins of the differential cross-section measurements are estimated by using a bootstrap procedure (as outlined in Section~\ref{sec:unfolding}) and included in the covariance matrix in the likelihood function. Each source of systematic uncertainty in the measurement is also included in the covariance matrix and is assumed to be fully correlated between the bins of \mfl{} and \mjj{}. Scale and PDF uncertainties in the  \sherpa{} strong \zzjj{}  and \mglo{} EW \zzjj{}  predictions are implemented as Gaussian-constrained nuisance parameters.
The confidence level at each value of Wilson coefficient is calculated using the profile-likelihood test statistic~\cite{Feldman:1997qc}, which is assumed to be distributed according to a $\chi^2$ distribution with one degree of freedom~\cite{wilks1938}.
From this, 95\% confidence intervals are constructed for each Wilson coefficient. 
The first bin of the \mfl{} distribution is not used in the statistical test. This bin is insensitive to the dimension-eight operators, but removing the bin prevents a possible overconstraint on the Wilson coefficients, which arises due to one very small eigenvalue of the combined \mfl{} -- \mjj{} covariance matrix.
 
\begin{table}[t]
\centering
\begin{tabular}{cccc}
\hline \hline
Wilson        &  $|\mathcal{M}_{\mathrm{d8}}|^2$   & \multicolumn{2}{c}{95\% confidence interval [\TeV{}$^{-4}$]} \\
coefficient  &  Included      &   Expected    &   Observed  \\
\hline
$f_{\textrm{T},0}/\Lambda^4$ & yes & [-1.00, 0.97] &[-0.98, 0.93]  \\
& no & [-19, 19] & [-23, 17] \\
\hline
$f_{\textrm{T},1}/\Lambda^4$ & yes & [-1.3, 1.3] & [-1.2, 1.2]  \\
& no & [-140, 140] & [-160, 120]   \\
\hline
$f_{\textrm{T},2}/\Lambda^4$ & yes  & [-2.6, 2.5] & [-2.5, 2.4] \\
& no & [-63, 62] & [-74, 56]  \\
\hline
$f_{\textrm{T},5}/\Lambda^4$ & yes & [-2.6, 2.5] &  [-2.5, 2.4]  \\
& no & [-68, 67] & [-79, 60]  \\
\hline
$f_{\textrm{T},6}/\Lambda^4$ & yes & [-4.1, 4.1] & [-3.9, 3.9] \\
& no & [-550, 540] & [-640, 480] \\
\hline
$f_{\textrm{T},7}/\Lambda^4$ & yes & [-8.8, 8.4] & [-8.5, 8.1]  \\
& no  & [-220, 220]  & [-260, 200]\\
\hline
$f_{\textrm{T},8}/\Lambda^4$ & yes & [-2.2, 2.2] & [-2.1, 2.1]  \\
& no & [-3.9, 3.8]$\times10^{4}$ & [-4.6, 3.1]$\times10^{4}$  \\
\hline
$f_{\textrm{T},9}/\Lambda^4$ & yes & [-4.7, 4.7] & [-4.5, 4.5]  \\
& no & [-6.4, 6.3]$\times10^{4}$ & [-7.5, 5.5]$\times10^{4}$  \\
\hline \hline
\end{tabular}
\vspace{0.5cm}
\caption{Expected and observed 95\% confidence interval for the dimension-eight Wilson coefficients, using a two-dimensional fit to the \zzjj{} differential cross-sections measured as a function of \mfl{} and \mjj{}. Results are presented when including or excluding the pure dimension-eight contributions to the EFT prediction.}
\label{tab:dim8}
\end{table}
 
\begin{figure}[h]
\includegraphics[width=0.49\textwidth]{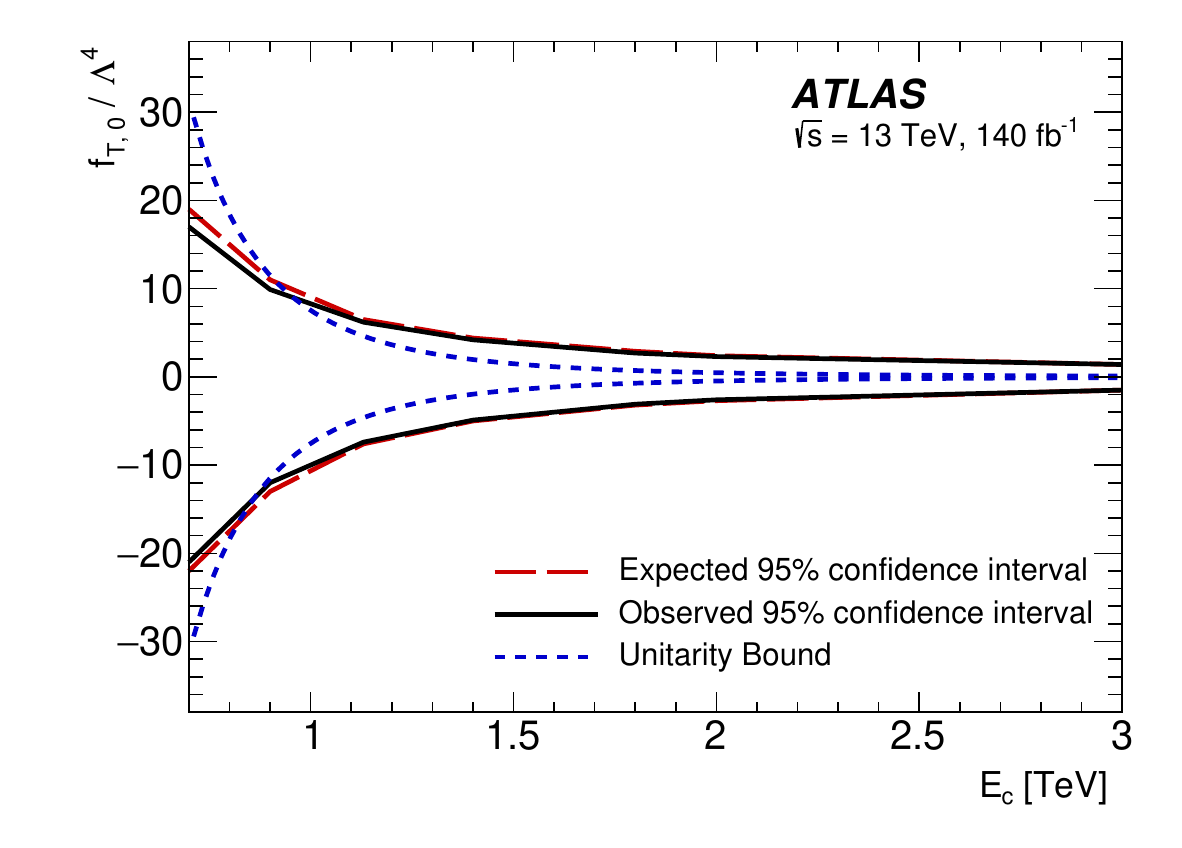}
\includegraphics[width=0.49\textwidth]{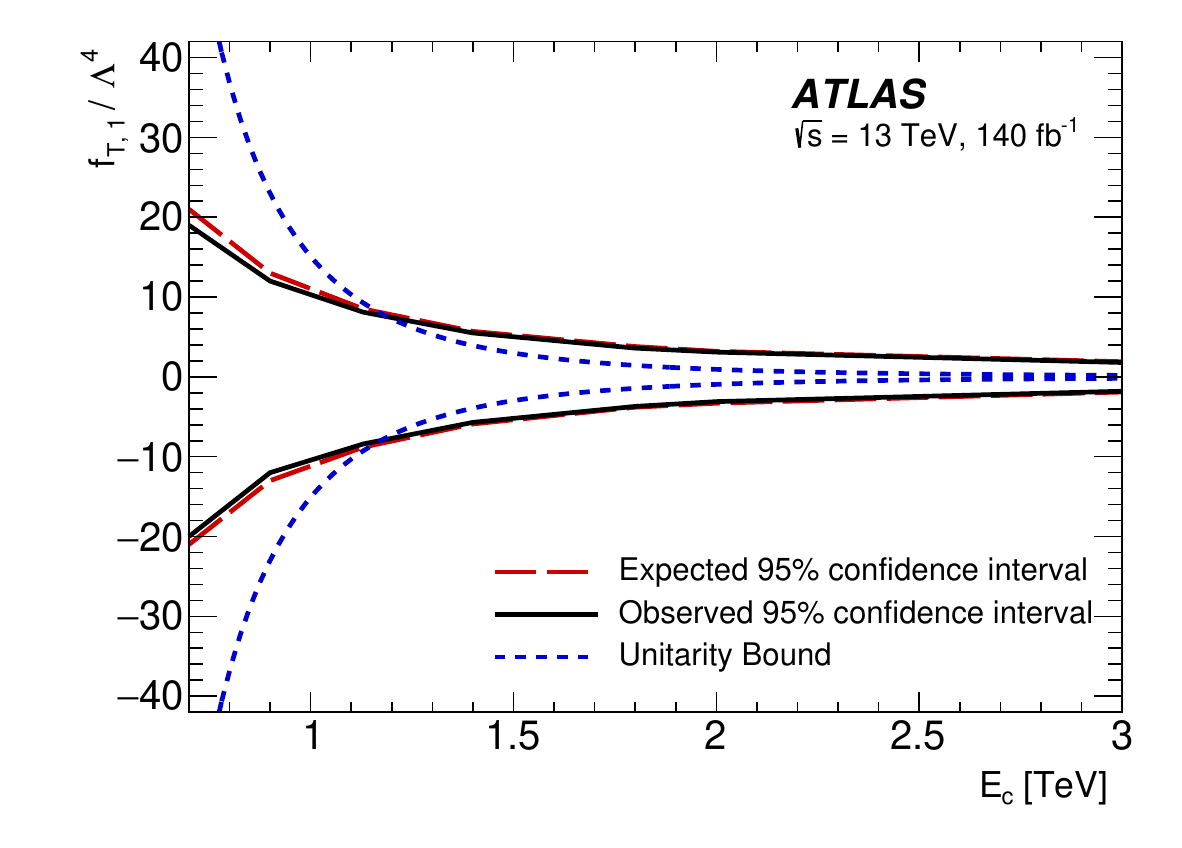}
 
\caption{Expected and observed 95\% confidence interval for the $f_{\textrm{T,0}}$ and $f_{\textrm{T,1}}$ Wilson coefficients as a function of a cut-off scale, $E_c$, which
restricts the interference- and pure dimension-eight- contributions to have $\mfl<E_c$. The constraints are obtained using a two-dimensional fit to the \zzjj{} differential cross-sections measured as a function of \mfl{} and \mjj{}.
\label{fig:clipping}}
\end{figure}

The 95\% confidence intervals on the Wilson coefficients in the dimension-eight effective field theory are shown in Table~\ref{tab:dim8}.
For each Wilson coefficient, confidence intervals are shown when including or excluding the pure dimension-eight contribution in the theoretical prediction. In all cases, the Wilson coefficients are consistent with zero. The Wilson coefficients associated with the $\mathcal{O}_{\textrm{T,0}}$ and $\mathcal{O}_{\textrm{T,1}}$ operators are the most tightly constrained. Furthermore, the constraints obtained with the pure dimension-eight contribution included in the theoretical prediction are more stringent than those obtained with only the interference contribution included in the theoretical calculation. This means that the limits are only valid if higher-order terms in the EFT expansion do not contribute significantly.
\FloatBarrier
 
Constraints are also placed on each Wilson coefficient after restricting the interference- and pure dimension-eight- contributions to have $\mfl<E_c$, where $E_c$ is a cut-off that prevents unitarity being violated at large energy scales. The dependence of the 95\% confidence intervals on the value of $E_c$ is shown in Figure~\ref{fig:clipping} for the $\mathcal{O}_{\textrm{T,0}}$ and $\mathcal{O}_{\textrm{T,1}}$ operators. The 95\% confidence intervals degrade by a factor of 4--5 when the energy scale cut off is reduced from $E_c=\infty$ to $E_c=1$~\TeV. A similar trend is seen for the other operators and shown in the Appendix. Unitarity bounds are also shown in Figure ~\ref{fig:clipping}, to indicate the consistency between the 95\% confidence intervals and the partial-wave unitarity considerations for vector boson scattering processes~\cite{Almeida:2020ylr,}.


\section{Conclusion}
\label{sec:conclusion}

Differential cross-sections are measured for the production of four charged leptons in association with two jets, using proton--proton collision data collected by the ATLAS experiment at a centre-of-mass energy of $\sqrt{s}=13$~\TeV{} and with an integrated luminosity of 140~fb$^{-1}$. This final state is sensitive to the strong \zzjj{} process, in which the jets arise from the strong interactions, and  the electroweak \zzjj{} process, which is characterised by the $t$-channel exchange of an electroweak boson.  The measurement of \zzjj{}  production can therefore be used to improve  the understanding of both the electroweak and strong interactions that underpin the Standard Model of particle physics.
The differential cross-sections for \zzjj{}  production are measured 
as a function of observables that collectively (i) characterise vector-boson scattering processes, (ii) probe the polarisation, parity and charge conjugation properties of the \zzjj{}  process, and (iii) probe the real emission of quarks and gluons from the \zzjj{}  process. 
The differential cross-sections are compared with various state-of-the-art Monte Carlo event generator predictions and the  measurements are found to be sensitive to the  event-generator modelling of strong \zzjj{} production and the EW \zzjj{} process at high dijet invariant mass and high values of $|\Dphijj{}|$. The differential cross-section measurements are consistent with Standard Model expectations and are used to set constraints on anomalous weak-boson self-interactions induced by dimension-six and dimension-eight operators in Standard Model effective field theory.


\section*{Acknowledgements}


We thank CERN for the very successful operation of the LHC, as well as the
support staff from our institutions without whom ATLAS could not be
operated efficiently.
 
We acknowledge the support of
ANPCyT, Argentina;
YerPhI, Armenia;
ARC, Australia;
BMWFW and FWF, Austria;
ANAS, Azerbaijan;
CNPq and FAPESP, Brazil;
NSERC, NRC and CFI, Canada;
CERN;
ANID, Chile;
CAS, MOST and NSFC, China;
Minciencias, Colombia;
MEYS CR, Czech Republic;
DNRF and DNSRC, Denmark;
IN2P3-CNRS and CEA-DRF/IRFU, France;
SRNSFG, Georgia;
BMBF, HGF and MPG, Germany;
GSRI, Greece;
RGC and Hong Kong SAR, China;
ISF and Benoziyo Center, Israel;
INFN, Italy;
MEXT and JSPS, Japan;
CNRST, Morocco;
NWO, Netherlands;
RCN, Norway;
MEiN, Poland;
FCT, Portugal;
MNE/IFA, Romania;
MESTD, Serbia;
MSSR, Slovakia;
ARRS and MIZ\v{S}, Slovenia;
DSI/NRF, South Africa;
MICINN, Spain;
SRC and Wallenberg Foundation, Sweden;
SERI, SNSF and Cantons of Bern and Geneva, Switzerland;
MOST, Taiwan;
TENMAK, T\"urkiye;
STFC, United Kingdom;
DOE and NSF, United States of America.
In addition, individual groups and members have received support from
BCKDF, CANARIE, Compute Canada and CRC, Canada;
PRIMUS 21/SCI/017 and UNCE SCI/013, Czech Republic;
COST, ERC, ERDF, Horizon 2020 and Marie Sk{\l}odowska-Curie Actions, European Union;
Investissements d'Avenir Labex, Investissements d'Avenir Idex and ANR, France;
DFG and AvH Foundation, Germany;
Herakleitos, Thales and Aristeia programmes co-financed by EU-ESF and the Greek NSRF, Greece;
BSF-NSF and MINERVA, Israel;
Norwegian Financial Mechanism 2014-2021, Norway;
NCN and NAWA, Poland;
La Caixa Banking Foundation, CERCA Programme Generalitat de Catalunya and PROMETEO and GenT Programmes Generalitat Valenciana, Spain;
G\"{o}ran Gustafssons Stiftelse, Sweden;
The Royal Society and Leverhulme Trust, United Kingdom.
 
The crucial computing support from all WLCG partners is acknowledged gratefully, in particular from CERN, the ATLAS Tier-1 facilities at TRIUMF (Canada), NDGF (Denmark, Norway, Sweden), CC-IN2P3 (France), KIT/GridKA (Germany), INFN-CNAF (Italy), NL-T1 (Netherlands), PIC (Spain), ASGC (Taiwan), RAL (UK) and BNL (USA), the Tier-2 facilities worldwide and large non-WLCG resource providers. Major contributors of computing resources are listed in Ref.~\cite{ATL-SOFT-PUB-2021-003}.


\clearpage
\appendix
\part*{Appendix}
\label{sec:appendixA}

The differential cross-sections for inclusive \zzjj{} production in the VBS-enhanced region as a function of \pTfl{}, \pTjj{}, \Dyjj{}, and \cosTStwo{} are shown in Figure \ref{fig:unfolded-vbsenh2}. The differential cross-sections for inclusive \zzjj{} production in the VBS-suppressed region as a function of \pTfl{}, \pTjj{}, \pTfljj{}, $S_{\textrm{T}, 4 \ell jj}$, \Dyjj{}, \Dphijj{}, \cosTS{} and \cosTStwo{} are shown in Figures \ref{fig:unfolded-vbssup2} and \ref{fig:unfolded-vbssup3}. The data are compared with two theoretical predictions, constructed from \mglo{} for the EW \zzjj{}  process and either \sherpa{} or \mgnlo{} for the strong \zzjj{}  process. The EW \zzjj{}  contribution is also shown separately to show the sensitivity of the measurements to the electroweak process.

\begin{figure}[tp]
\includegraphics[width=0.5\textwidth]{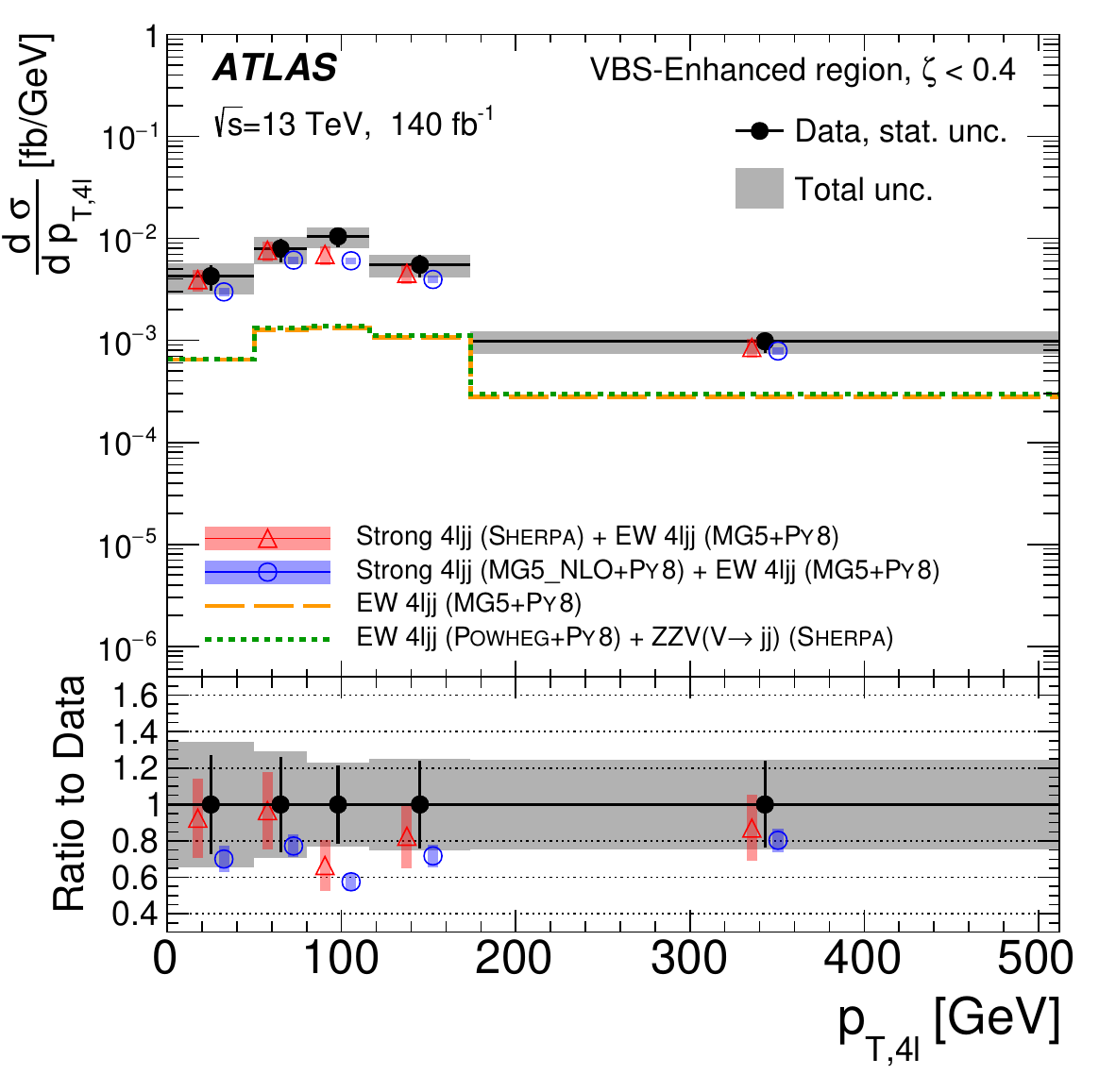}
\includegraphics[width=0.5\textwidth]{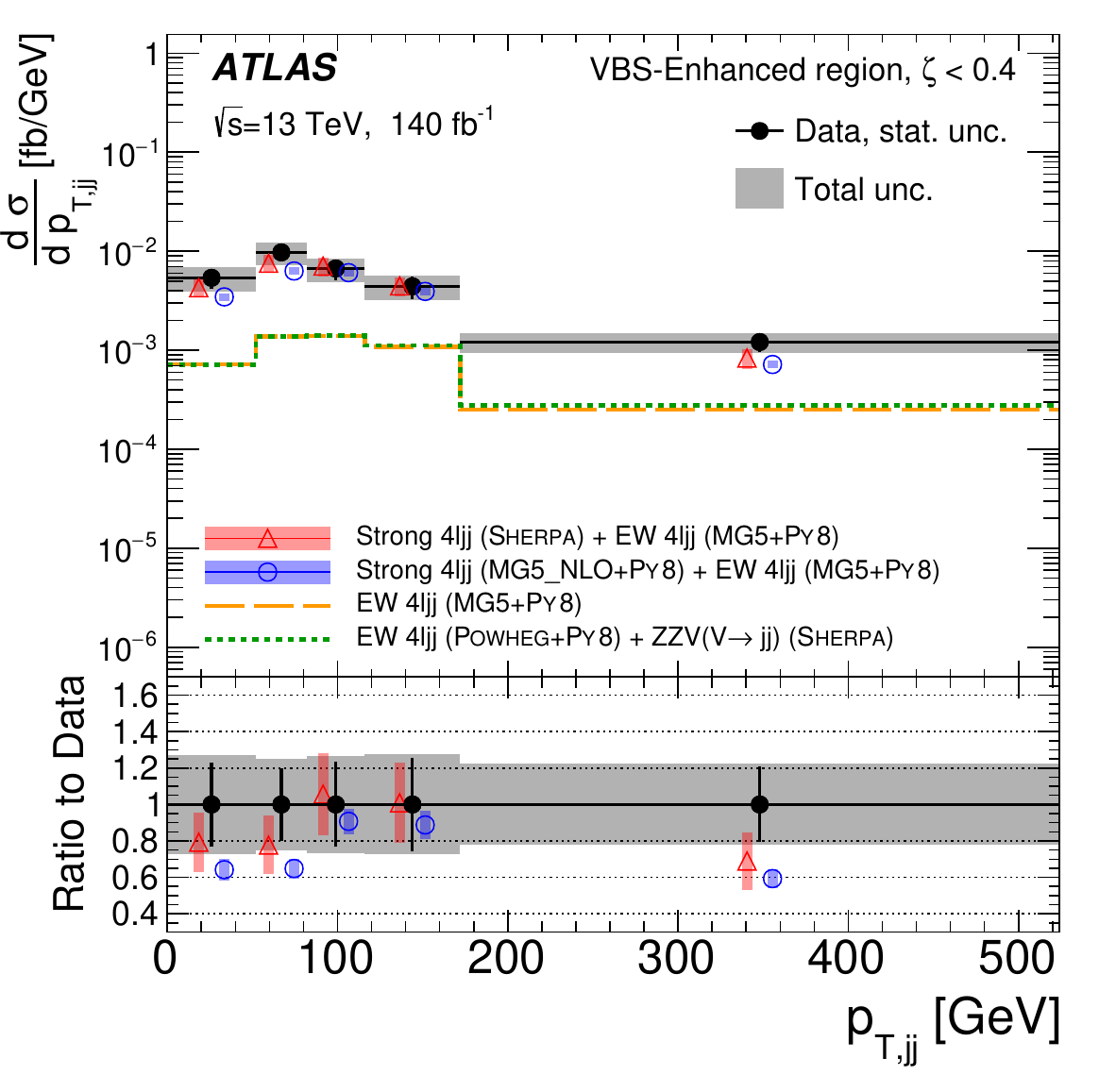}\\
\includegraphics[width=0.5\textwidth]{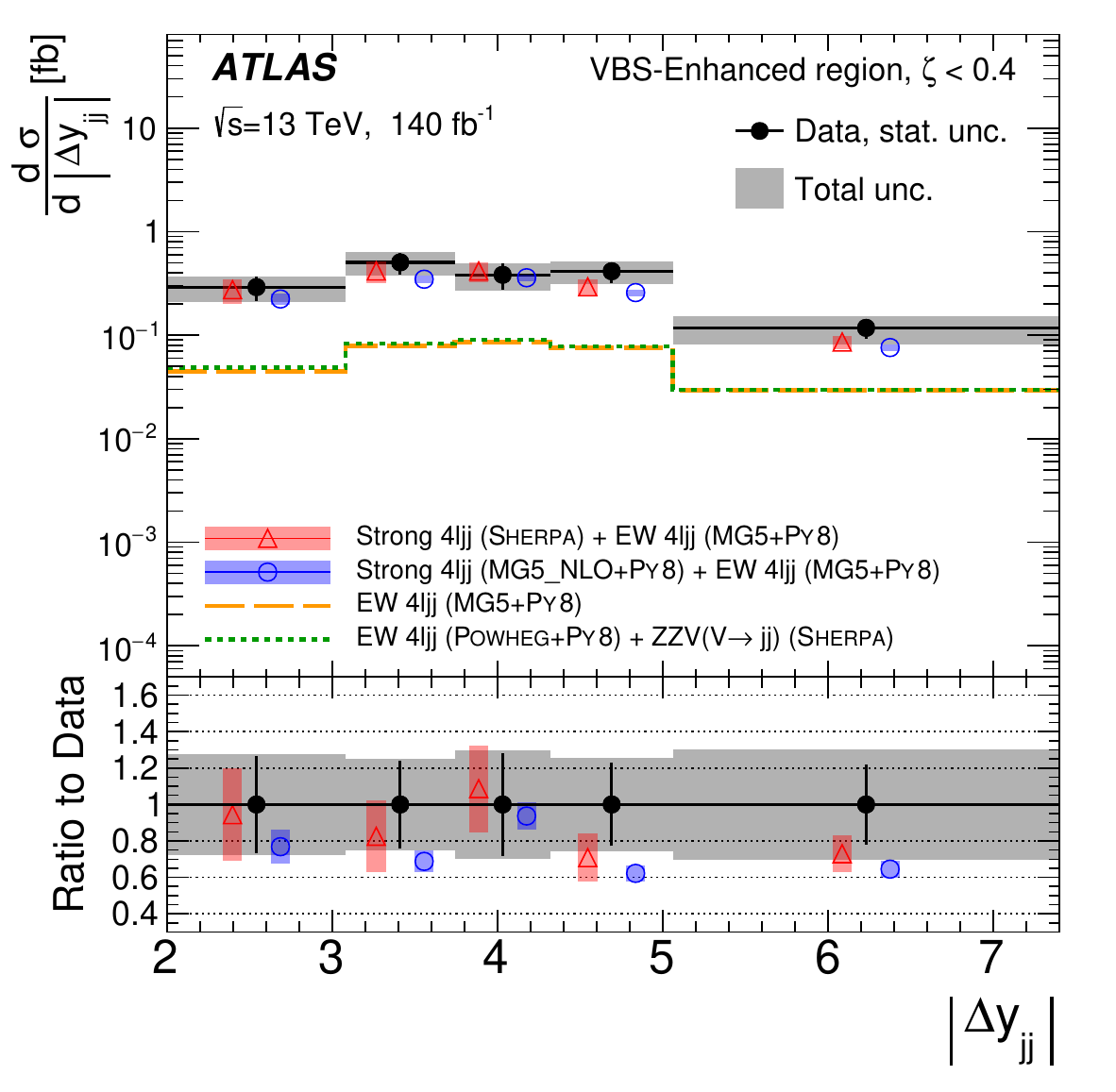}
\includegraphics[width=0.5\textwidth]{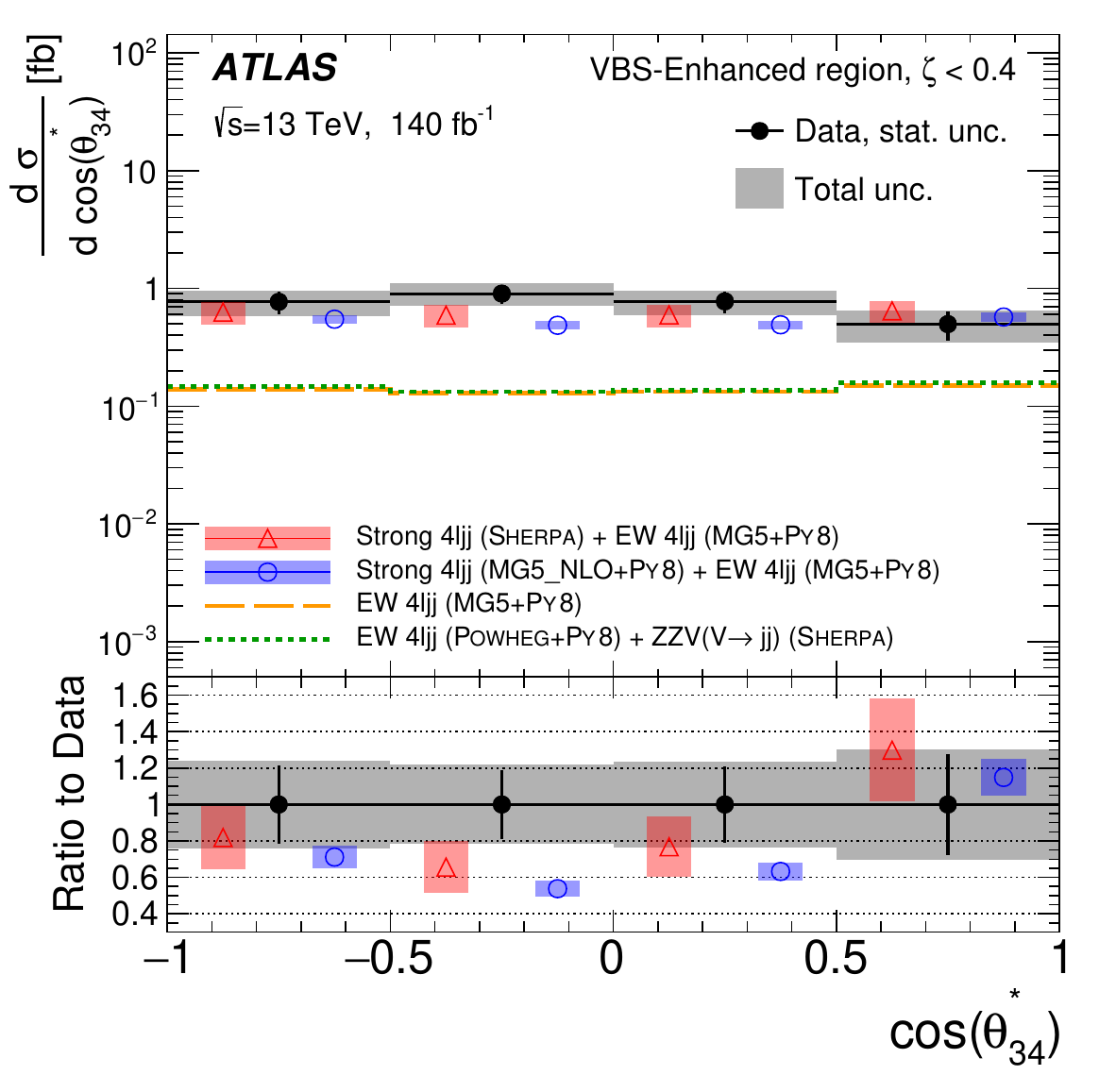}\\
 
\caption{Differential cross-sections for inclusive \zzjj{}  production in the VBS-enhanced region as a function of \pTfl{} (top left), \pTjj (top right),
\Dyjj{} (bottom left) and \cosTStwo{} for the second $Z$ boson candidate (bottom right).
The data are represented as black points and the associated error bars represent the statistical uncertainty. The total uncertainty in the measurement is represented as a grey hatched band. The data are compared to two theoretical predictions, estimated by using \sherpa{} (triangles) and \mg{} (circles) for the strong \zzjj{} contribution and \mglo{} for the EW \zzjj{}  contribution. The band on the theoretical predictions represents the theoretical uncertainty from renormalisation/factorisation scale choices and PDF choice. The dashed lines show the contribution of EW \zzjj{} production to the differential cross-section as predicted by \mglo{} and \powpyt{}. The $s$-channel contributions from $ZZV$ production are missing from the \powpyt{} prediction and are estimated with
\sherpa{}. For the  \pTfl{}, \pTjj and
\Dyjj{} measurements, any `overflow' events that lie above the upper bin edge of the last bin are included in that bin.
\label{fig:unfolded-vbsenh2}}
\end{figure}

\begin{figure}[tp]
\includegraphics[width=0.5\textwidth]{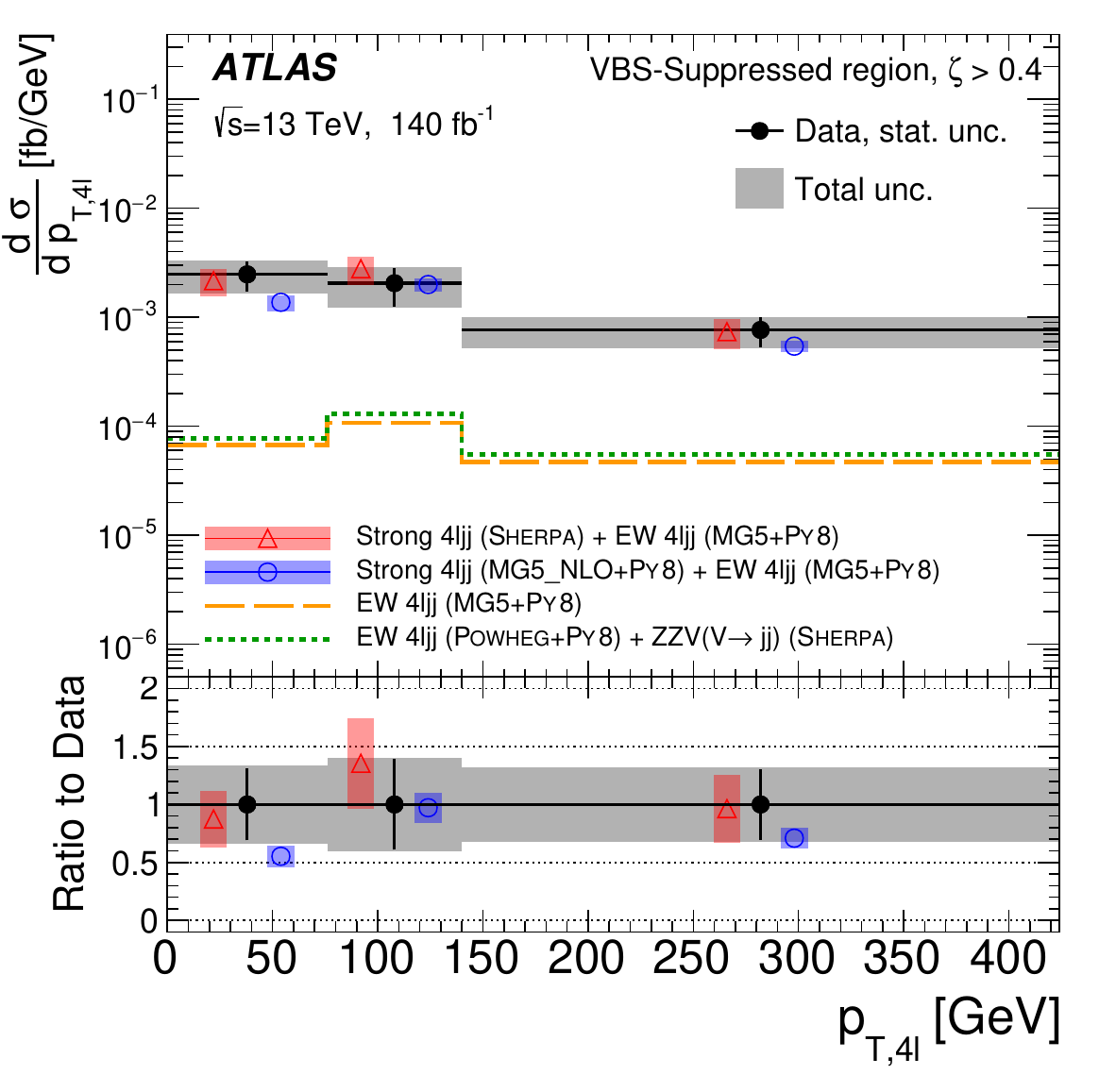}
\includegraphics[width=0.5\textwidth]{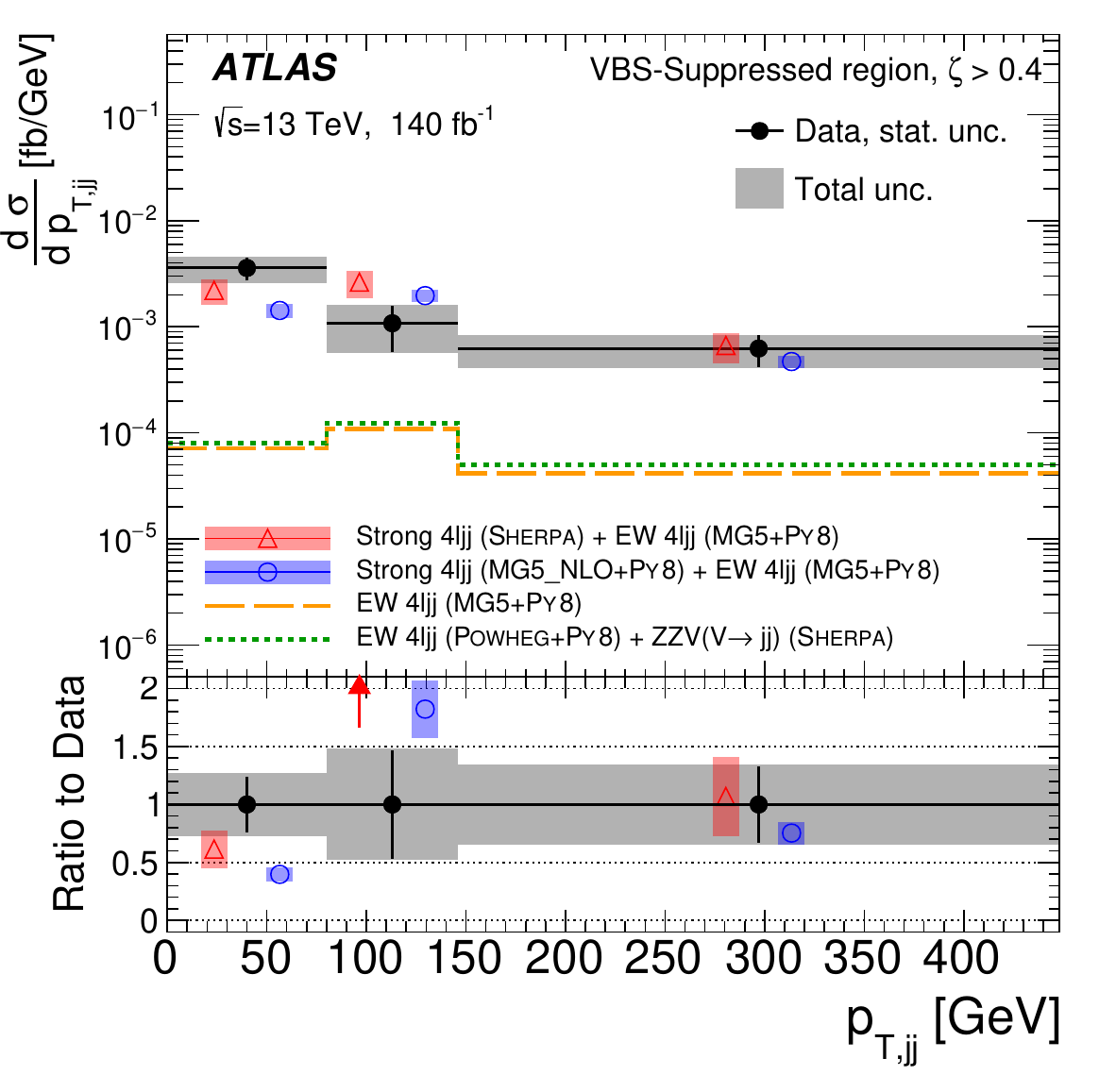}\\
\includegraphics[width=0.5\textwidth]{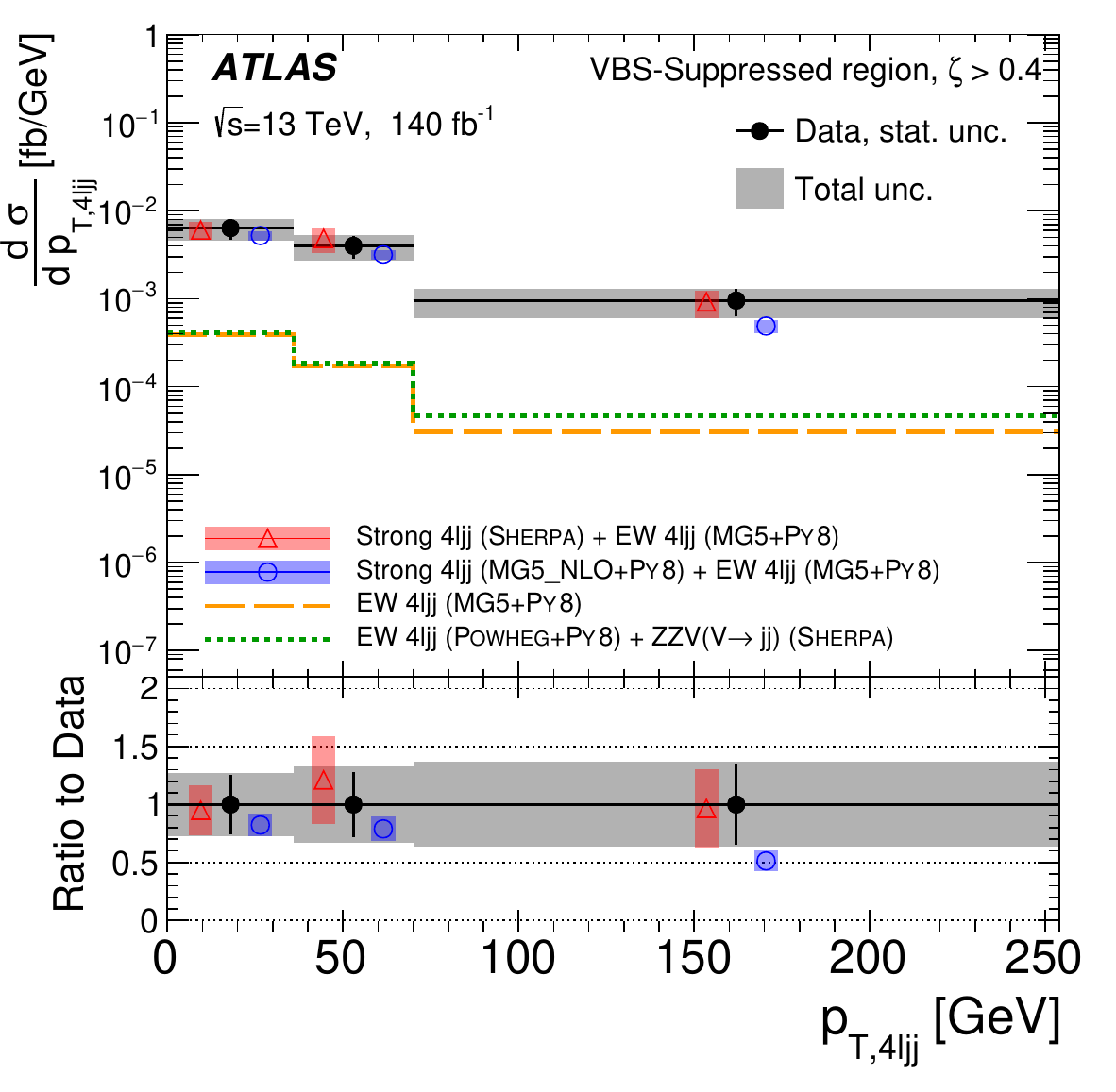}
\includegraphics[width=0.5\textwidth]{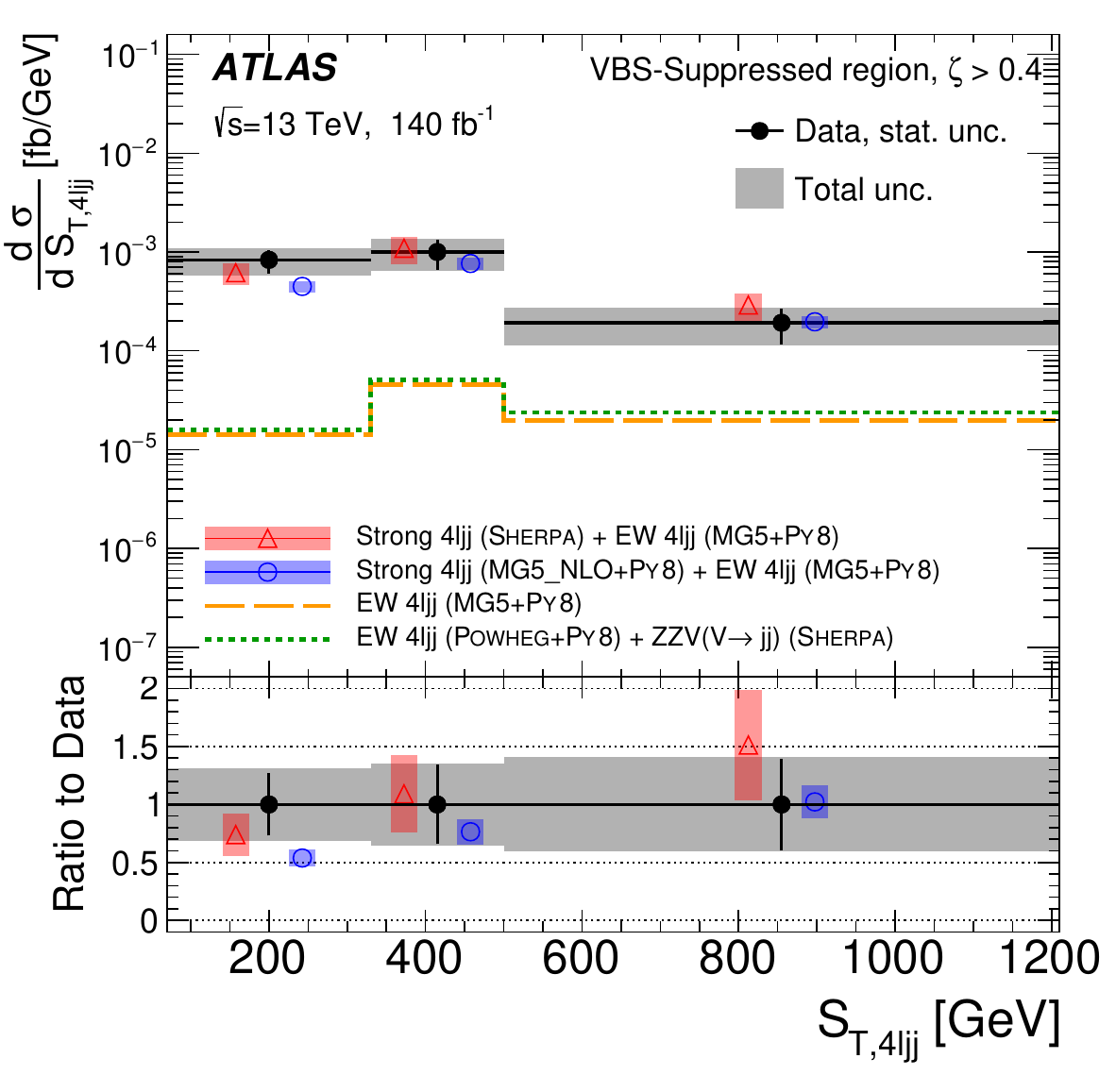}\\
\caption{Differential cross-sections for inclusive \zzjj{}  production in the VBS-suppressed region as a function of \pTfl{} (top left), \pTjj (top right),
\pTfljj{} (bottom left) and $S_{\textrm{T}, 4 \ell jj}$ (bottom right),
The data are represented as black points and the associated error bars represent the statistical uncertainty. The total uncertainty in the measurement is represented as a grey hatched band. The data are compared to two theoretical predictions, estimated using \sherpa{} (triangles) and \mg{} (circles) for the strong \zzjj{} contribution and \mglo{} for the EW \zzjj{}  contribution. The band on the theoretical predictions represents the theoretical uncertainty from renormalisation/factorisation scale choices and PDF choice. The dashed lines show the contribution of EW \zzjj{} production to the differential cross-section as predicted by \mglo{} and \powpyt{}. The $s$-channel contributions from $ZZV$ production are missing from the \powpyt{} prediction and are estimated with
\sherpa{}. `Overflow' events that lie above the upper bin edge of the last bin are included in that bin.
\label{fig:unfolded-vbssup2}}
\end{figure}
 
\begin{figure}[tp]
\includegraphics[width=0.5\textwidth]{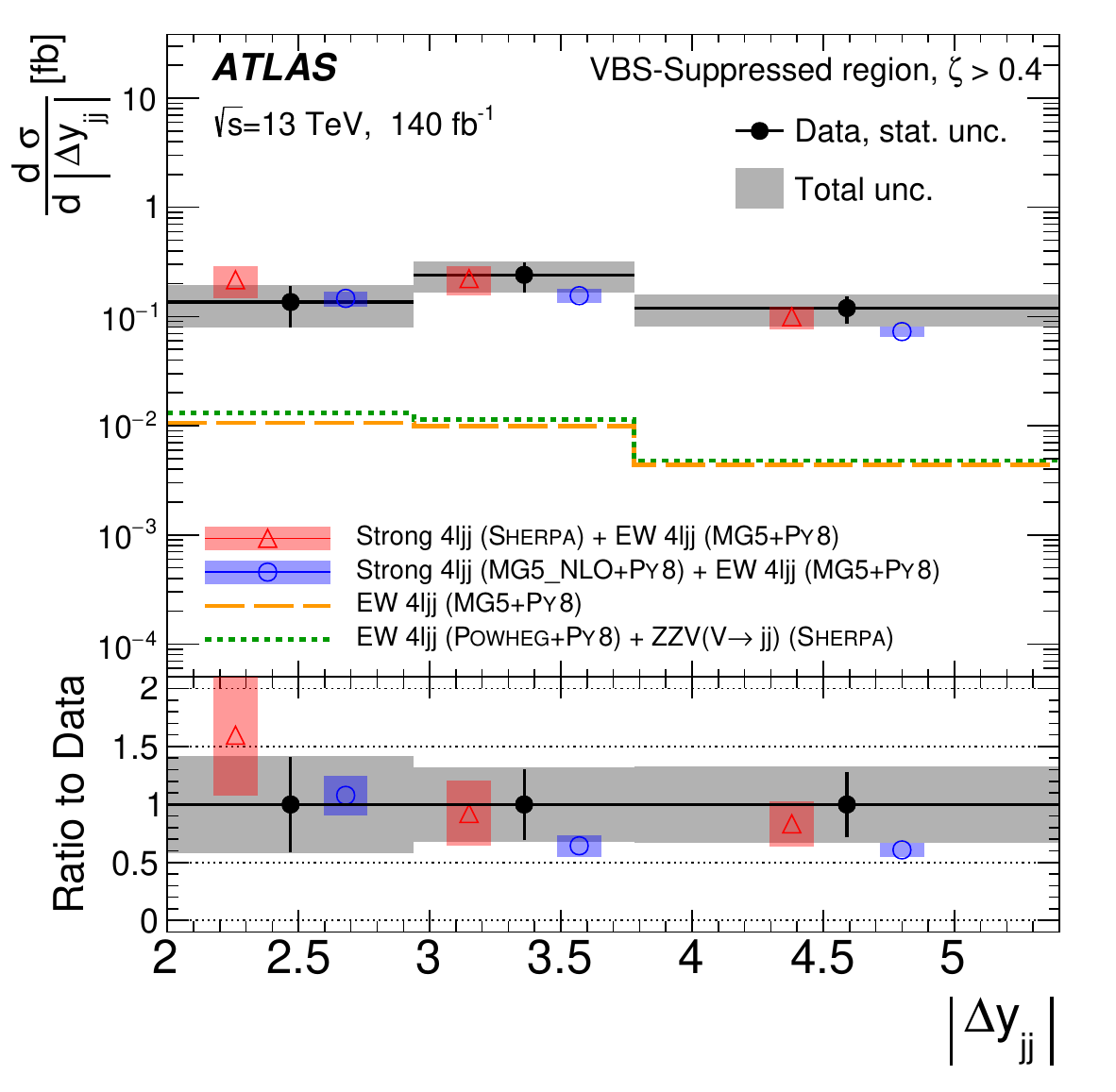}
\includegraphics[width=0.5\textwidth]{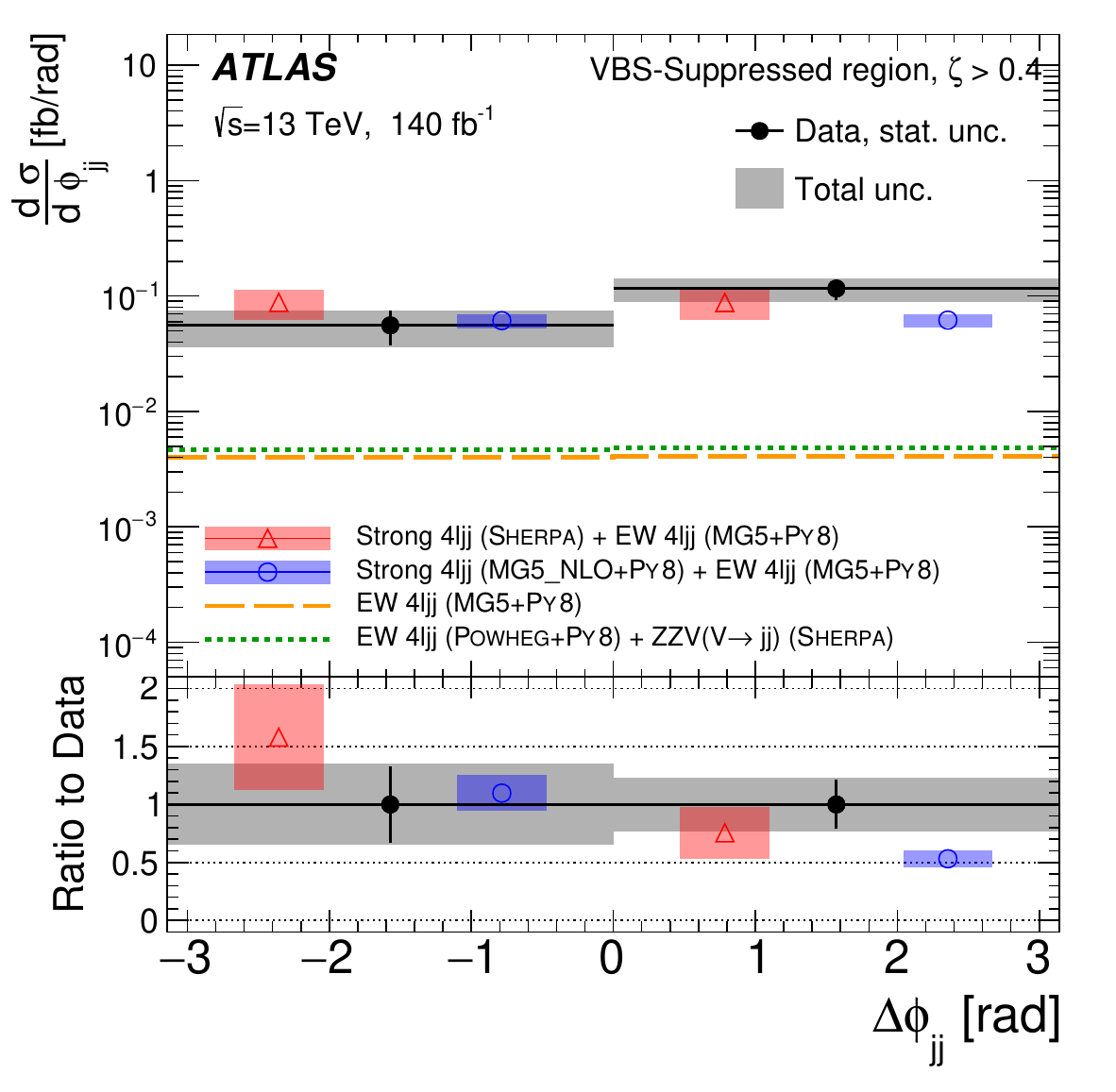}\\
\includegraphics[width=0.5\textwidth]{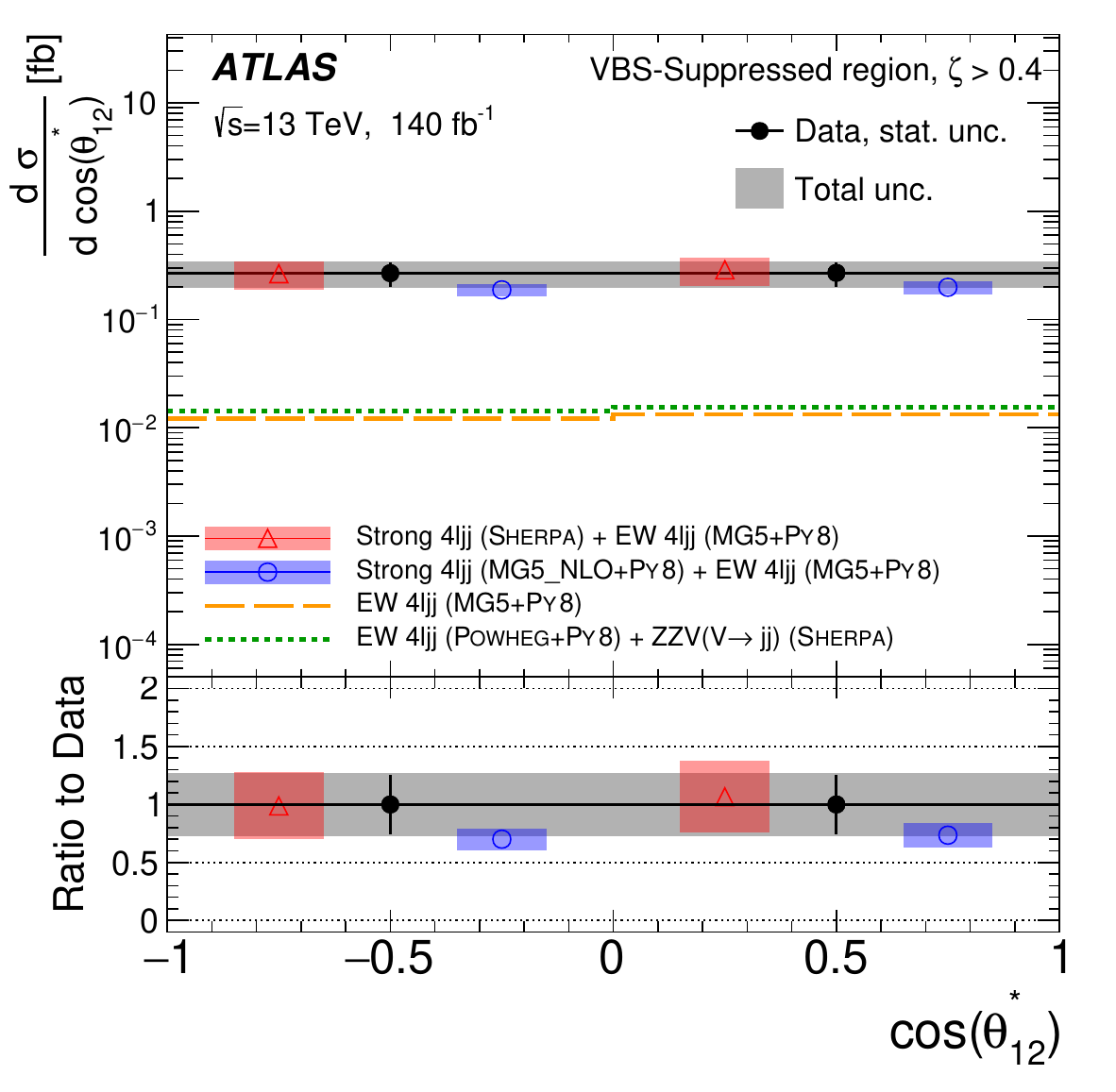}
\includegraphics[width=0.5\textwidth]{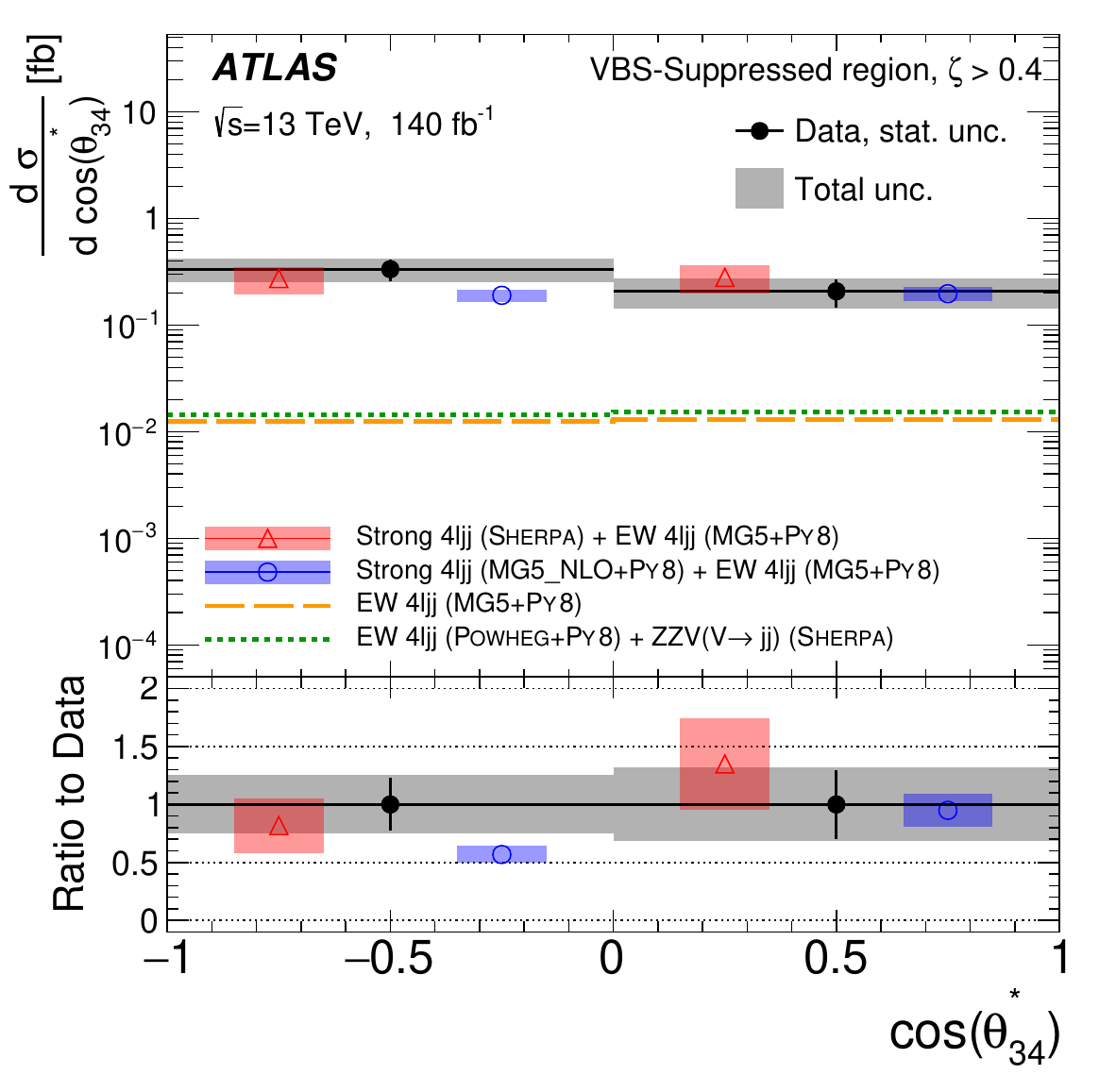}\\
 
\caption{Differential cross-sections for inclusive \zzjj{}  production in the VBS-suppressed region as a function of \Dyjj{} (top left), \Dphijj (top right),
\cosTS{} (bottom left) and \cosTStwo (bottom right),
The data are represented as black points and the associated error bars represent the statistical uncertainty. The total uncertainty in the measurement is represented as a grey hatched band. The data are compared to two theoretical predictions, estimated using \sherpa{} (triangles) and \mg{} (circles) for the strong \zzjj{} contribution and \mglo{} for the EW \zzjj{}  contribution. The band on the theoretical predictions represent the theoretical uncertainty from renormalisation/factorisation scale choices and PDF choice. The dashed lines show the contribution of EW \zzjj{} production to the differential cross-section as predicted by \mglo{} and \powpyt{}. The $s$-channel contributions from $ZZV$ production are missing from the \powpyt{} prediction and are estimated with
\sherpa{}.  For the \Dyjj{} measurement, any `overflow' events that lie above the upper bin edge of the last bin are included in that bin.
\label{fig:unfolded-vbssup3}}
\end{figure}
 
The dependence of the 95\% confidence intervals on the value of $E_c$ is shown in Figure~\ref{fig:clippingtwo} for the $\mathcal{O}_{\textrm{T,2}}$, $\mathcal{O}_{\textrm{T,5}}$, $\mathcal{O}_{\textrm{T,6}}$, $\mathcal{O}_{\textrm{T,7}}$, $\mathcal{O}_{\textrm{T,8}}$ and $\mathcal{O}_{\textrm{T,9}}$ operators. The 95\% confidence intervals degrade by a factor of 4--5 when the energy scale cut off is reduced from $E_c=\infty$ to $E_c=1$~\TeV.

\begin{figure}[h]
\includegraphics[width=0.46\textwidth]{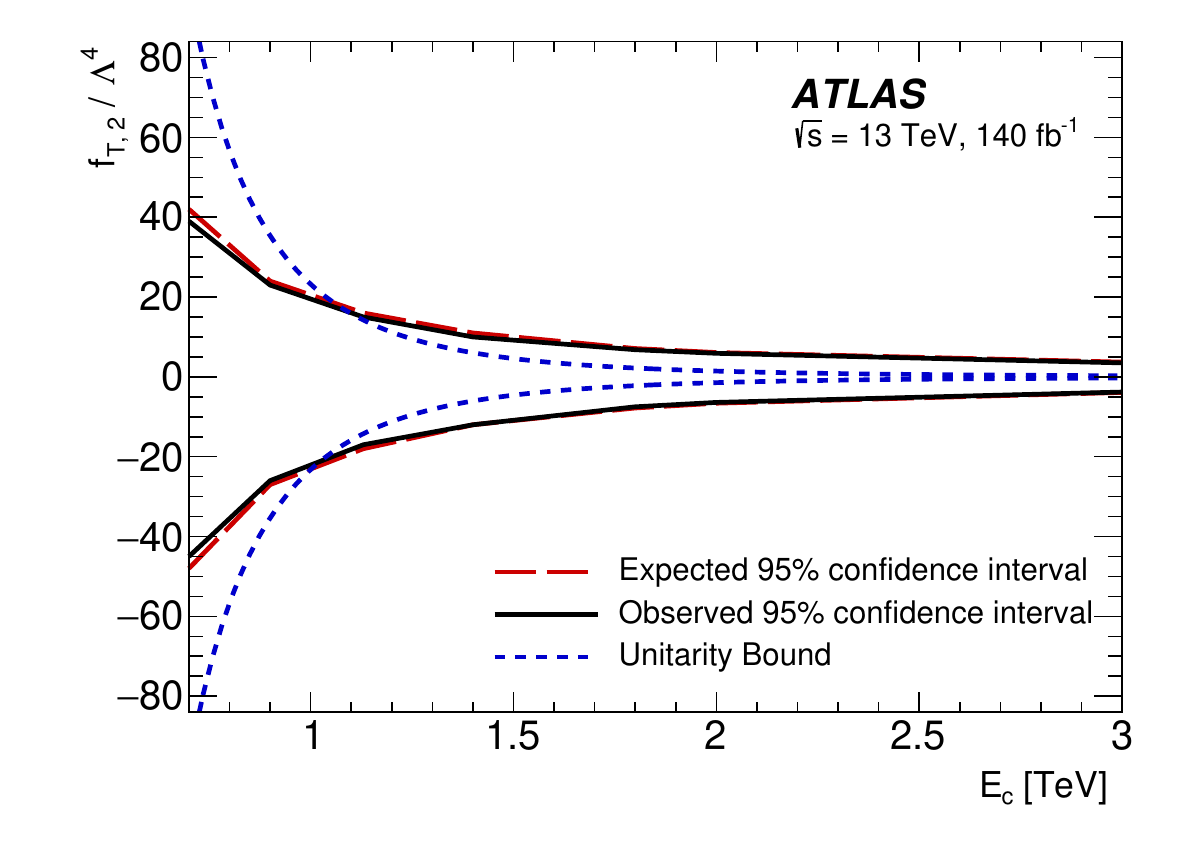}
\includegraphics[width=0.46\textwidth]{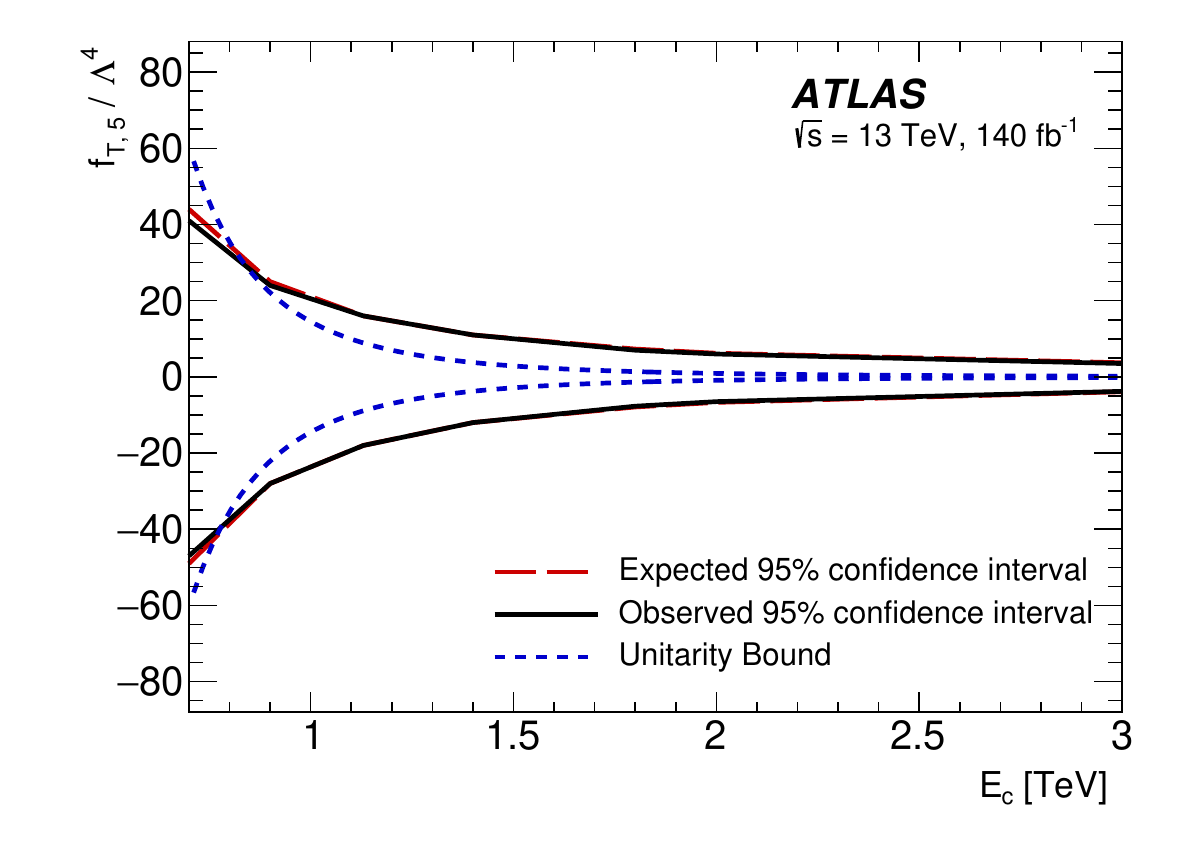}\\
\includegraphics[width=0.46\textwidth]{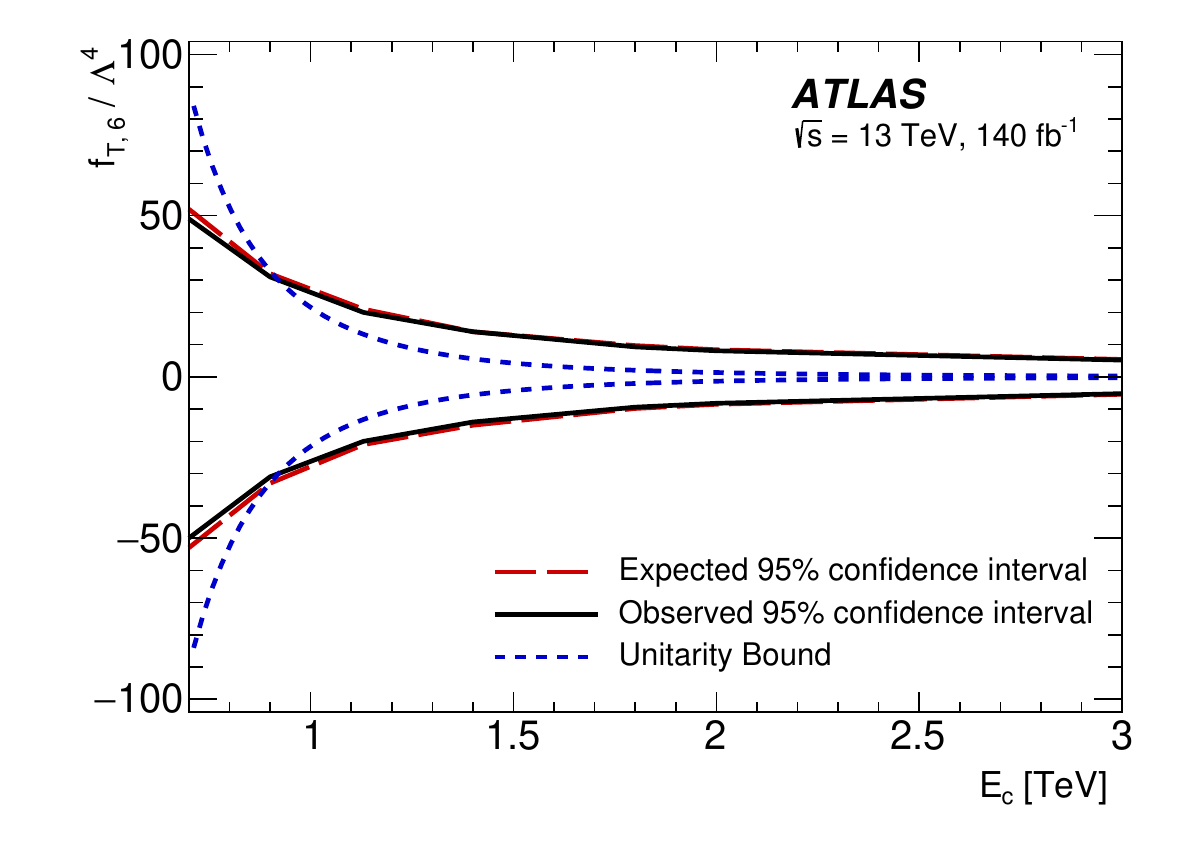}
\includegraphics[width=0.46\textwidth]{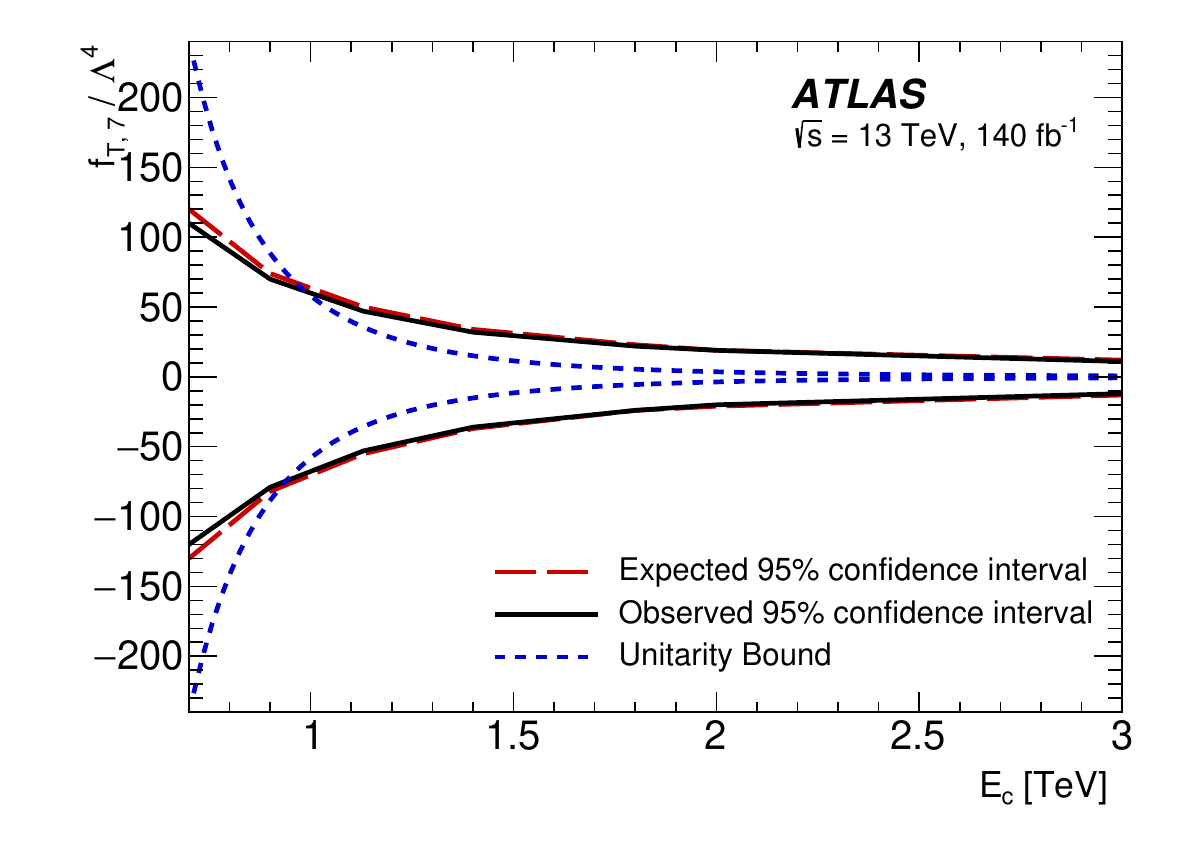}\\
\includegraphics[width=0.46\textwidth]{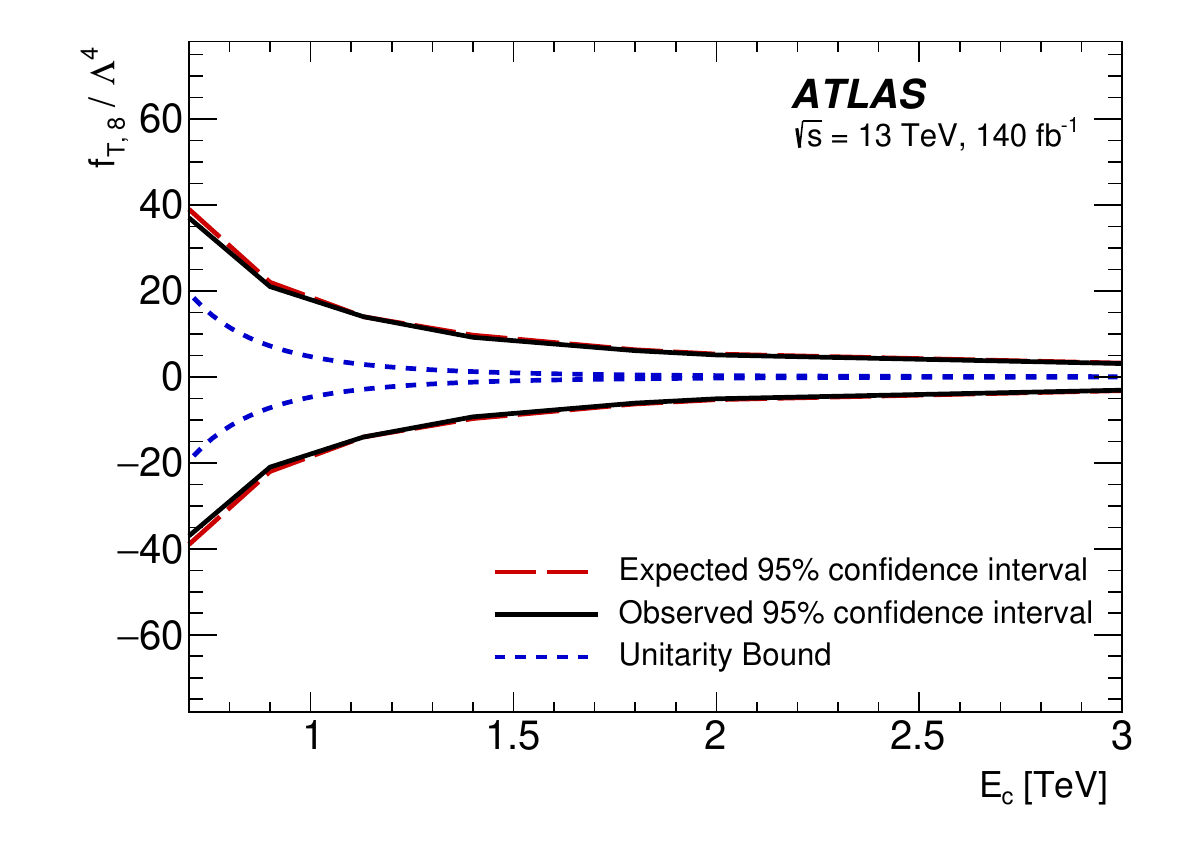}
\includegraphics[width=0.46\textwidth]{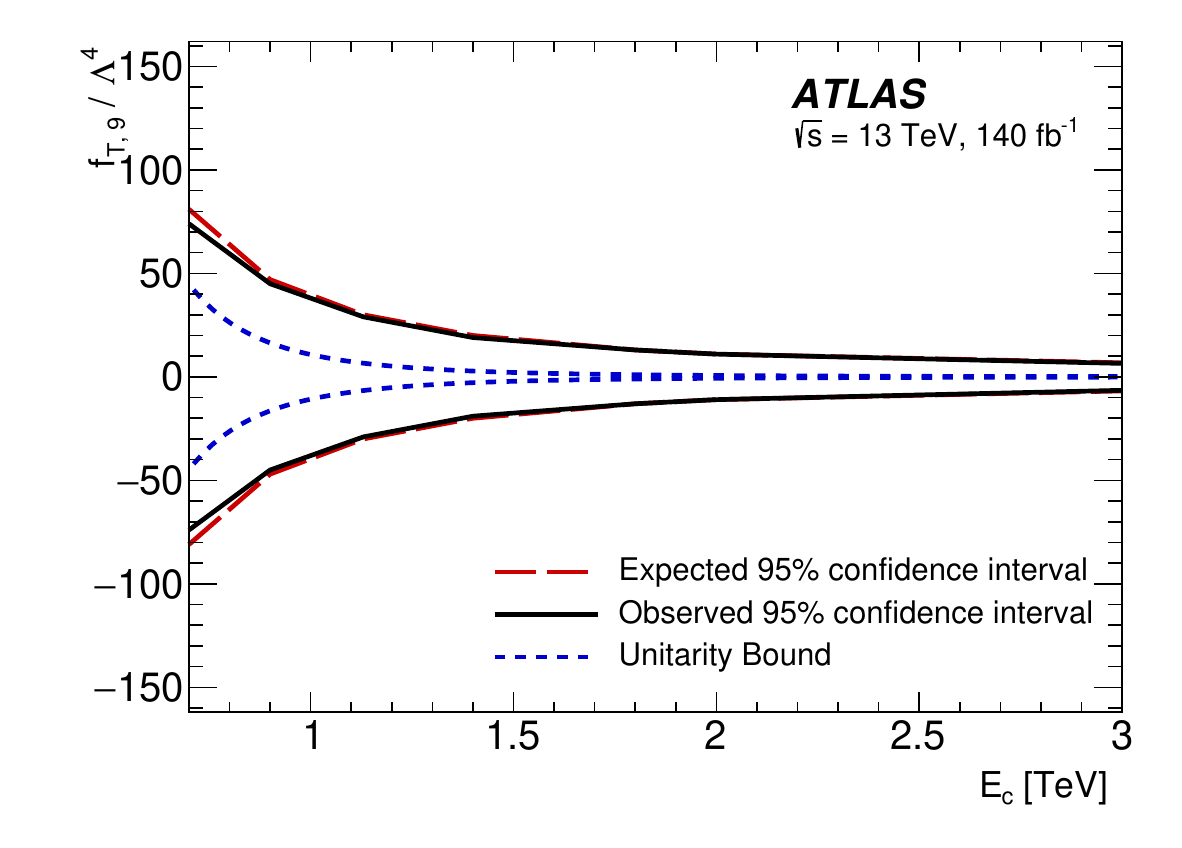}\\
 
\caption{Expected and observed 95\% confidence interval for the $f_{\textrm{T,2}}$,  $f_{\textrm{T,5}}$, $f_{\textrm{T,6}}$,  $f_{\textrm{T,7}}$, $f_{\textrm{T,8}}$ and $f_{\textrm{T,9}}$ Wilson coefficients as a function of a cut-off scale, $E_c$, which
restricts the interference- and pure dimension-eight- contributions to have $\mfl<E_c$. The constraints are obtained using a two-dimensional fit to the \zzjj{} differential cross-sections measured as a function of \mfl{} and \mjj{}.
\label{fig:clippingtwo}}
\end{figure}

Table \ref{tab:dim6} shows 95\% confidence intervals on the Wilson coefficients of operators in a dimension-six effective field theory~\cite{Brivio:2017btx}. The method to constrain the Wilson coefficients is identical to that used for the dimension-eight operators in Section~\ref{sec:eft},
with the interactions from the dimension-six operators provided by the SMEFTSim package \cite{Brivio:2017btx}. In the case of CP-odd operators, the interference contribution is zero for parity-even observables such as \mjj{} and \mfl{}. However, the interference contribution produces large asymmetric effects in the parity-odd \Dphijj{} observable. Constraints are therefore placed on the CP-odd Wilson coefficients using the measured differential cross-section as a function of \Dphijj{} when the pure dimension-six contribution to the EFT is excluded. The constraints obtained on the dimension-six Wilson coefficients in this analysis are much weaker than those obtained measurements of diboson production~\cite{ATLAS:2019rob,ATLAS:2021kog,CMS:2020mxy, CMS:2021icx, CMS:2021cxr} and vector-boson fusion~\cite{ATLAS:2020nzk}.
 
\begin{table}[bp]
\centering
\begin{tabular}{cccc}
\hline \hline
Wilson        &  $|\mathcal{M}_{\mathrm{d6}}|^2$   & \multicolumn{2}{c}{95\% confidence interval [\TeV{}$^{-2}$]} \\
coefficient  &  Included      &   Expected    &   Observed  \\
\hline
$c_{{W}}/\Lambda^2$ & yes & [-1.3, 1.3] & [-1.2, 1.2] \\
& no & [-32, 32] & [-37, 28] \\
\hline
$c_{\widetilde{W}}/\Lambda^2$ & yes& [-1.3, 1.3] & [-1.2, 1.2] \\
& no & [-17, 17]$^{*}$ & [0, 30]$^{*}$ \\
\hline
$c_{{HWB}}/\Lambda^2$ & yes& [-16, 7] & [-16, 6] \\
& no & [-12, 12] & [-15, 10] \\
\hline
$c_{{H\widetilde{W}B}}/\Lambda^2$ & yes &  [-1.3, 1.3] & [-1.2, 1.2] \\
& no & [-67, 67]$^{*}$ & [-25, 130]$^{*}$ \\
\hline
$c_{{HB}}/\Lambda^2$ & yes& [-13, 13] & [-12, 12] \\
& no & [-38, 38] & [-38, 38] \\
\hline
$c_{{H\widetilde{B}}}/\Lambda^2$ & yes &  [-13, 13] & [-12, 12] \\
& no & [-420, 420]$^{*}$ & [-200, 790]$^{*}$ \\
\hline
\hline
\vspace{0.3cm}
\end{tabular}
\caption{Expected and observed 95\% confidence interval for the dimension-six Wilson coefficients. Results are presented when including or excluding the pure dimension-six contributions to the EFT prediction. The constraints are obtained using a two-dimensional fit to the \zzjj{} differential cross-sections measured as a function of \mfl{} and \mjj{}, except for the constraints on CP-odd operators when the pure dimension-six contribution to the EFT is excluded. Those constraints, denoted by a ($^*$), are obtained in a fit to the differential cross-section as a function of \Dphijj{}.}
\label{tab:dim6}
\end{table}


\addcontentsline{toc}{part}{Appendix}
\clearpage

\printbibliography
 
 
\begin{flushleft}
\hypersetup{urlcolor=black}
{\Large The ATLAS Collaboration}

\bigskip

\AtlasOrcid[0000-0002-6665-4934]{G.~Aad}$^\textrm{\scriptsize 102}$,
\AtlasOrcid[0000-0002-5888-2734]{B.~Abbott}$^\textrm{\scriptsize 120}$,
\AtlasOrcid[0000-0002-1002-1652]{K.~Abeling}$^\textrm{\scriptsize 55}$,
\AtlasOrcid[0000-0001-5763-2760]{N.J.~Abicht}$^\textrm{\scriptsize 49}$,
\AtlasOrcid[0000-0002-8496-9294]{S.H.~Abidi}$^\textrm{\scriptsize 29}$,
\AtlasOrcid[0000-0002-9987-2292]{A.~Aboulhorma}$^\textrm{\scriptsize 35e}$,
\AtlasOrcid[0000-0001-5329-6640]{H.~Abramowicz}$^\textrm{\scriptsize 151}$,
\AtlasOrcid[0000-0002-1599-2896]{H.~Abreu}$^\textrm{\scriptsize 150}$,
\AtlasOrcid[0000-0003-0403-3697]{Y.~Abulaiti}$^\textrm{\scriptsize 117}$,
\AtlasOrcid[0000-0003-0762-7204]{A.C.~Abusleme~Hoffman}$^\textrm{\scriptsize 137a}$
% ...
\bigskip
\\

% ...
$^{29}$Physics Department, Brookhaven National Laboratory, Upton NY; United States of America.\\
% ...
$^{35}$$^{(a)}$Facult\'e des Sciences Ain Chock, R\'eseau Universitaire de Physique des Hautes Energies - Universit\'e Hassan II, Casablanca;$^{(b)}$Facult\'{e} des Sciences, Universit\'{e} Ibn-Tofail, K\'{e}nitra;$^{(c)}$Facult\'e des Sciences Semlalia, Universit\'e Cadi Ayyad, LPHEA-Marrakech;$^{(d)}$LPMR, Facult\'e des Sciences, Universit\'e Mohamed Premier, Oujda;$^{(e)}$Facult\'e des sciences, Universit\'e Mohammed V, Rabat;$^{(f)}$Institute of Applied Physics, Mohammed VI Polytechnic University, Ben Guerir; Morocco.\\
% ...
$^{49}$Fakult\"{a}t Physik , Technische Universit{\"a}t Dortmund, Dortmund; Germany.\\
% ...
$^{55}$II. Physikalisches Institut, Georg-August-Universit\"{a}t G\"ottingen, G\"ottingen; Germany.\\
% ...
$^{102}$CPPM, Aix-Marseille Universit\'e, CNRS/IN2P3, Marseille; France.\\
% ...
$^{117}$Department of Physics, New York University, New York NY; United States of America.\\
% ...
$^{120}$Homer L. Dodge Department of Physics and Astronomy, University of Oklahoma, Norman OK; United States of America.\\
% ...
$^{137}$$^{(a)}$Departamento de F\'isica, Pontificia Universidad Cat\'olica de Chile, Santiago;$^{(b)}$Millennium Institute for Subatomic physics at high energy frontier (SAPHIR), Santiago;$^{(c)}$Instituto de Investigaci\'on Multidisciplinario en Ciencia y Tecnolog\'ia, y Departamento de F\'isica, Universidad de La Serena;$^{(d)}$Universidad Andres Bello, Department of Physics, Santiago;$^{(e)}$Instituto de Alta Investigaci\'on, Universidad de Tarapac\'a, Arica;$^{(f)}$Departamento de F\'isica, Universidad T\'ecnica Federico Santa Mar\'ia, Valpara\'iso; Chile.\\
% ...
$^{150}$Department of Physics, Technion, Israel Institute of Technology, Haifa; Israel.\\
$^{151}$Raymond and Beverly Sackler School of Physics and Astronomy, Tel Aviv University, Tel Aviv; Israel.\\
% ...
$^{a}$ Also Affiliated with an institute covered by a cooperation agreement with CERN.\\
$^{b}$ Also at An-Najah National University, Nablus; Palestine.\\
$^{c}$ Also at Borough of Manhattan Community College, City University of New York, New York NY; United States of America.\\
$^{d}$ Also at Center for High Energy Physics, Peking University; China.\\
$^{e}$ Also at Center for Interdisciplinary Research and Innovation (CIRI-AUTH), Thessaloniki; Greece.\\
$^{f}$ Also at Centro Studi e Ricerche Enrico Fermi; Italy.\\
% ...

\end{flushleft}


\end{document}